\documentclass[fleqn,usenatbib]{mnras}

\usepackage{graphicx,times}
\usepackage{subfigure}
\newcommand{\be}{\begin{equation}}
\usepackage{threeparttable}
\usepackage{booktabs}
\newcommand{\ee}{\end{equation}}
\newcommand{\bea}{\begin{eqnarray}}
\newcommand{\eea}{\end{eqnarray}}

\usepackage{amsmath}
\usepackage{cases}
\usepackage{longtable}
\usepackage{hyperref}
\usepackage{epstopdf}
\usepackage{amsmath,bm}
\usepackage{amssymb}
\usepackage{natbib}
\usepackage{morefloats}
\usepackage{multirow}
\usepackage{array}
\usepackage{verbatim}
\usepackage{color}
\usepackage{lineno}

\usepackage[normalem]{ulem} 
\usepackage{bm} 
\usepackage[varvw]{newtxmath}  

\newcommand{\cs} {c_{\rm s}}
\newcommand{\gamef}{\gamma_{\rm eff}}
\newcommand{\LJ} {L_{\rm J}}
\newcommand{\Ma} {M_{\rm A}}
\newcommand{\pcc} {\hbox{ cm}^{-3}}
\newcommand{\tff} {t_{\rm ff}}
\newcommand{\va} {v_{\rm A}}
\newcommand{\vc} {v_{\rm c}}

\newcommand{\vg} {v_{\rm g}}
\newcommand{\vgvec} {{\bm v}_{\rm g}}
\newcommand{\vs} {v_{\rm s}}
\newcommand{\vsvec} {{\bm v}_{\rm s}}
\newcommand{\vturb} {v_{\rm trb}}
\newcommand{\vturbvec} {{\bm v}_{\rm trb}}

\title[Properties of collapse-driven MHD turbulence]{On the properties and implications of collapse-driven MHD turbulence}

\author[V\'azquez-Semadeni et al.]{
Enrique V\'azquez-Semadeni,$^1$\thanks{E-mail:e.vazquez@irya.unam.mx}
Yue Hu,$^2$
Siyao Xu,$^3$
Rub\'en Guerrero-Gamboa$^1$
and Alex Lazarian$^2$
\\
$^1$Instituto de Radioastronom\'ia y Astrof\'isica, Universidad Nacional Aut\'onoma de M\'exico, Apdo.\ Postal 3-72, Morelia, Michoac\'an, 58090, M\'exico\\
$^2$Department of Astronomy, University of Wisconsin-Madison, Madison, WI 53706, USA\\
$^3$Department of Physics, University of Florida, 2001 Museum Rd., Gainesville, FL 32611, USA
}

\begin{document}
\label{firstpage}
\pagerange{\pageref{firstpage}--\pageref{lastpage}}
\maketitle


\begin{abstract}
We investigate the driving of MHD turbulence by gravitational contraction using simulations of an idealized, initially spherical, isothermal, globally magnetically supercritical  molecular cloud core in the presence of initial transonic and trans-Alfv\'enic turbulence. To this end, we perform a Helmholtz decomposition of the velocity field, and investigate the evolution of its solenoidal and compressible parts, as well as of the velocity component along the gravitational acceleration vector, which can be considered as the infall component of the velocity field. We find that: 1) In spite of being supercritical, the core first contracts to a sheet perpendicular to the mean magnetic field, and the sheet itself collapses. 2) The solenoidal component of the turbulence remains at roughly its initial level throughout the of the simulation, while the compressible component increases continuously. This implies that turbulence does {\it not} dissipate towards the center of the core. 3) The distribution of simulation cells in the $B$-$\rho$ plane occupies a wide triangular region at low densities, bounded below by the expected trend for fast MHD waves ($B \propto \rho$, applicable for high local Alfv\'enic Mach number $\Ma$) and above by the trend expected for slow waves ($B \sim$ constant, applicable for low local $\Ma$). At high densities, the distribution follows a single trend $B \propto \rho^{\gamef}$, with $1/2 < \gamef < 2/3$, as expected for gravitational compression.  4) The mass-to-magnetic flux ratio $\lambda$ increases with the radius $r$ out to which it is measured in the core, due to the different scalings of the mass and magnetic flux with $r$. At a fixed radius, $\lambda$ increases with time due to the accretion of material along field lines.  5) The solenoidal energy fraction is much smaller than the total turbulent component, indicating that the collapse drives the turbulence mainly compressibly, even in directions orthogonal to that of the collapse.

\end{abstract}

\begin{keywords}
MHD -- Gravitation -- Turbulence -- ISM: clouds -- stars: formation 
\end{keywords}

\section{Introduction} \label{sec:intro}

In recent years, the driving of turbulence by gravitational collapse at various scales has received considerable attention, in particular in relation to whether enough gravitational energy is available in the collapsing material for driving the turbulence in the central accreting objects, from the scale of accreting galactic disks to molecular clouds to protostellar disks \citep{Klessen_Henneb10}; whether it can act as a possible reservoir for the gravitational energy released during the collapse, so that this energy could be stored in the turbulence and possibly delay the collapse \citep[e.g.,] [] {RobG12, Murray_Chang15, Murray+17, Li18, Xu_Laz20}, and what is its equivalent thermodynamic behavior \citep[e.g.,] [] {VS+98, GV20}.

However, one issue that has not been studied in depth is whether the random motions driven by collapse 
really qualify as turbulence, exhibiting standard turbulence properties. Indeed, the nature of the driving in the collapse-driven case is significantly different from that in other, more standard cases. For example, the energy-injection scale shrinks over time rather than being constant, at least during the prestellar stage of the collapse. In the particular case of the collapse of molecular cloud cores (objects of typical densities $n \sim 10^4 \pcc$ and sizes $\sim 0.1$ pc), this can be understood because the prestellar stage of collapse in spherical geometry is characterized by a flat-density central core with a radius of the order of the Jeans length, at which the largest infall speeds also occur \citep[e.g.,] [] {WS85, Keto_Caselli10, Naranjo+15}. Since the central density increases over time, the Jeans length for the central density decreases over time. Thus, the energy-injection scale decreases over time, if it is of the order of the Jeans length, where the infall speed peaks.

In addition, if the energy-injection rate is of the order of the release rate of gravitational energy at the Jeans length, then it is also expected to vary over time, as it is given approximately by \citep{GV20}
\begin{equation}
 \dot E_{\rm g} \approx - \left( \frac{2|E_{\rm g}|^3} {M_{L_{\rm J}} L_{\rm J}^2}\right)^{1/2},
\label{eq:dotEg}
\end{equation}
where $L_{\rm J}$ is the Jeans length at the central density, $M_{L_{\rm J}}$ is the mass contained within a radius $R = L_{\rm J}$, and $E_{\rm g} \approx GM^2(L_{\rm J})/L_{\rm J}$ is the gravitational energy of this mass distribution. Therefore, since $L_{\rm J}$ decreases over time, both $E_{\rm g}$ and $\dot E_{\rm g}$ increase (in absolute value) over time (since $M_{\rm J}$ is expected to be constant). In summary, both the energy-injection scale and the energy-injection rate vary over time during the collapse, thus calling for an examination of whether the turbulence driven by gravitational contraction maintains the properties of turbulence driven at a fixed rate and scale. Indeed, \citet{GV20} found that, for the collapsing case, the turbulent energy appears to approach a ``pseudo-virial'' state, in which the kinetic energy is approximately half the gravitational energy, even though the system is far from equilibrium and both energies are increasing in time.

In this paper, we, therefore, examine numerically some of the main features of the MHD turbulence that develops during the prestellar stage of the gravitational collapse of an initially spherical core, with transonic and trans-Alfv\'enic initial velocity perturbations, employing a Helmholtz decomposition for the velocity field into its solenoidal and compressible parts. The former corresponds to the turbulence exclusively, while the latter contains the infall plus the turbulent components. In Sec.\ \ref{sec:num_sim} we describe our numerical simulation; in Sec.\ \ref{sec:method} we describe our analysis strategy. Then, in Sec.\ \ref{sec:results} we describe our main results, while in Sec.\ \ref{sec:disc} we discuss the interpretation and some implications of our results. Finally, in Sec.\ \ref{sec:concls} we present our conclusions.

\section{Numerical simulations} \label{sec:num_sim}

We perform and analyze three 3D numerical simulations of the {\it prestellar} stage of the collapse (i.e., before a singularity --- protostar --- forms) using the adaptive mesh refinement (AMR) code \textsc{FLASH}2.5 \citep{Fryxell+00}. The numerical simulations consist of an initial Gaussian density profile\footnote{Note that the choice of initial density profile is probably not important for the later evolution of the collapse, since the asymptotic spherical collapse solutions of \citet{WS85} have a well-defined Bonnor-Ebert-like profile, although the solutions are fully dynamical, and furthermore \citet{Gomez+21} have shown that an $r^{-2}$ density profile is an attractor for the profile's logarithmic slope in its outer power-law part, implying that the spherically-averaged density profile will spontaneously approach this slope as the collapse proceeds, as indeed observed in simulations with uniform \citep[e.g.,] [] {Larson69} or gaussian initial conditions \citep[e.g.,] [] {Naranjo+15}. Therefore, our gaussian initial profile can be considered as a ``generic'' initial condition, representative of a random turbulent density fluctuation.} embedded in a background of uniform density $\rho_0$, where the peak of the Gaussian is $2.5 \rho_0$ and the mean density of the box is $\langle\rho\rangle = 1.535 \rho_0$. The simulations are isothermal, and the density $\rho_0$ and sound speed $\cs$ are set so that the box length $L_\mathrm{0} \approx 2.5 \LJ$, where $\LJ=(\pi \cs^2 / G \langle \rho \rangle)^{1/2}$ is the Jeans length corresponding to the mean density in the numerical box.

Two of the simulations have the same setup and resolution, except that one of them is the purely hydrodynamic (HD) simulation {\it turb\_08} from \cite{GV20}, while the other ({\it mhdturb\_08}) is a magnetohydrodynamic (MHD) simulation, with the numerical box permeated by a uniform magnetic field oriented along the $z$-direction. The field strength $B_0$ is set so that $\cs = \va$, where $\va$ is the Alfv\'en speed. This condition results in the choice $B_0=\cs(4 \pi \langle\rho\rangle)^{1/2}$. The third simulation ({\it mhdturb\_10}) is identical to ({\it mhdturb\_08}), except that it has two additional levels of refinement, and is used to test for convergence in Appendix \ref{sec:convergence}.

Since the simulations are isothermal, we can rescale them using any set of values for which the box contains the same number of Jeans lengths and the MHD run satisfies the condition $\cs = \va$. For reference, we take fiducial physical values for the density, sound speed, and magnetic field (in the magnetized runs) of $n_0 = 4.86\times10^5\mathrm{\ cm}^{-3}$, $\cs = 0.21 \mathrm{\ km\ s}^{-1}$, and $B_0 = 106\ \mu G$, respectively, and the simulation size is $L_0 = 0.1$ pc per side. These values imply that the mass-to-flux ratio for the whole numerical box in the magnetic simulations {\it mhdturb\_08} and {\it mhdturb\_10}, normalized to the critical value, is $\lambda \approx 6$, so that these runs are strongly magnetically supercritical (i.e., not supported by the magnetic field strength).

For the unit of time, we choose the free-fall time $\tff$ for the mean density of the central Jeans mass in the box at the initial condition. This is computed as the mean density out to a radius where the mass internal to it equals the Jeans mass corresponding to the mean density out to that radius. 

The boundary conditions are periodic for the hydrodynamics, and isolated for the self-gravity. In runs {\it turb\_08} and {\it mhdturb\_08} we use a maximum refinement level of $\ell = 8$, corresponding to a maximum resolution of $2^{\ell+2} = 1024$ grid cells, of size $\approx 10^{-4}$ pc, or $\approx 20$ AU. We refine according to the Jeans per Jeans length in the HD simulation and 32 for the MHD simulation. The latter value follows the recommendation of \citet{Sur10}, of using at least 30 cells per Jeans length in an MHD simulation, in order to properly resolve the small-scale dynamo. Since increasing the number of cells per Jeans length in practice requires increasing the maximum resolution level, in Appendix \ref{sec:convergence} we check that our results do not vary significantly when increasing the maximum level to $\ell = 10$ in run {\it mhdturb\_10}.

With these refinement conditions, we can compute the highest density that is adequately resolved with the combination of the number of cells per Jeans length and the maximum refinement level \citep[eq.\ (32) from ][]{Federrath+10},
\begin{equation}
    \label{eq:dens_res}
    \rho_\mathrm{res} = \frac{\pi c_{s}^{2}}{4 G r_{\rm acc}^{2}} = \frac{\pi c_\mathrm{s}^{2}}{4 G } \left( \frac{ 2^{2+\ell} }{ j_\mathrm{r} L_0 } \right)^2 ,
\end{equation}
where $G$ is the gravitational constant, $\ell$ is the maximum refinement level, and $j_\mathrm{r}$ is the number of cells to resolve the Jeans length. In Table\ \ref{table:run_par} we summarize the resolution parameters for the various runs. The density is expressed as a number density, considering $n=\rho/\mu m_\mathrm{H}$, where $\mu$ is the mean molecular weight, which for molecular gas takes the value $\mu \approx 2.3$.  

Regions in the simulations with densities larger than $n_{\rm res}$ are expected to be affected by numerical dissipation.
Moreover, for densities $n \gtrsim 10^{10} \pcc$ the isothermal assumption may break down, as these densities correspond to the formation of a first hydrostatic core \citep[e.g.,] [] {Larson69}. In any case, as seen in Fig.\ \ref{fig:1}, these regions are very small, and failure to fully resolve them or to use a harder equation of state is unlikely to affect the global results we discuss here. On the other hand, the effects of numerical dissipation may indeed affect the results at very high densities, such on the $B$-$\rho$ correlation, although in this case, they may provide an emulation of the effects of ambipolar or reconnection diffusion processes, as we discuss in Sec.\ \ref{sec:B-n}.
\begin{table*}
\begin{center}
\caption{Simulation parameters.}
\begin{tabular}{lccc}
\hline
\hline
\label{table:run_par}
    Run         & Effective refinement & Cells per Jeans length & Maximum resolved density\\ 
                & $\Delta x$ [pc]      &   $j_{\rm r}$          & $n_\mathrm{res}$ [$\mathrm{cm}^{-3}$] \\   
    \hline
    turb\_08    & $9.76\times10^{-5}$       & 12	            & $1.091\times10^{8}$ \\
    mhdturb\_08 & $9.76\times10^{-5}$       & 32	            & $1.534\times10^{7}$ \\
    mhdturb\_10 & $2.44\times10^{-5}$       & 32	            & $2.455\times10^{8}$ \\
    \hline
\end{tabular}
\end{center}
\end{table*}

We first start the simulations with self-gravity turned off, and stir the gas with a turbulent forcing module \citep{PF10} for roughly one crossing time to introduce perturbations in the velocity, density and magnetic fields. The driving is fully solenoidal, and the energy is injected in a range of scales between 1/8 and 1/32 of the box. The turbulence generated reaches a transonic value of $\sigma \approx 0.8 \cs$, consistent with the typically observed turbulence levels at the size scale of the simulation \citep[$\sim 0.1$ pc; e.g.] [] {Heyer_Brunt04} Then, we turn off the forcing and turn on the self-gravity at $t=0$. Collapse ensues immediately, although initially, it is very slow compared to the turbulent motions. However, the turbulence decays during the early stages of the evolution, due to the absence of driving. Thus, during these stages of the collapse, the infall speed increases, while the turbulent velocity dispersion decreases, until the infall motions become strong enough to inject energy into the turbulent motions \citep[see] [for details] {GV20}.

\section{Methodology} \label{sec:method}

\subsection{Helmholtz decomposition} \label{ssec:Hel}

In this work, we adopt the Helmholtz theorem used in \citet{2022ApJ...941..133H} to decompose the velocity field into a solenoidal component (i.e., Alfv\'en mode) $\pmb{v}_{\rm s}$ and compressive component (i.e., fast and slow modes) $\pmb{v}_{\rm c}$:
\begin{equation}
\begin{aligned}
    \pmb{v}=\pmb{v}_{\rm s}+\pmb{v}_{\rm c}.
\end{aligned}
\end{equation}
The solenoidal and compressive components satisfy divergence free ($\nabla \cdot \pmb{v}_{\rm s}=0$) and curl free ($\nabla \times \pmb{v}_{\rm c}=0$) conditions, respectively. Owing to the Helmholtz theorem, $\pmb{v}_{\rm c}$ stems from a scalar potential $\phi$, i.e., $\pmb{v}_{\rm c}=-\nabla\phi$, and $\pmb{v}_{\rm s}$ stems from a vector potential $\pmb{\Phi}$, i.e., $\pmb{v}_{\rm s}=\nabla\times\pmb{\Phi}$.

\begin{figure*}
	\includegraphics[width=1.0\linewidth]{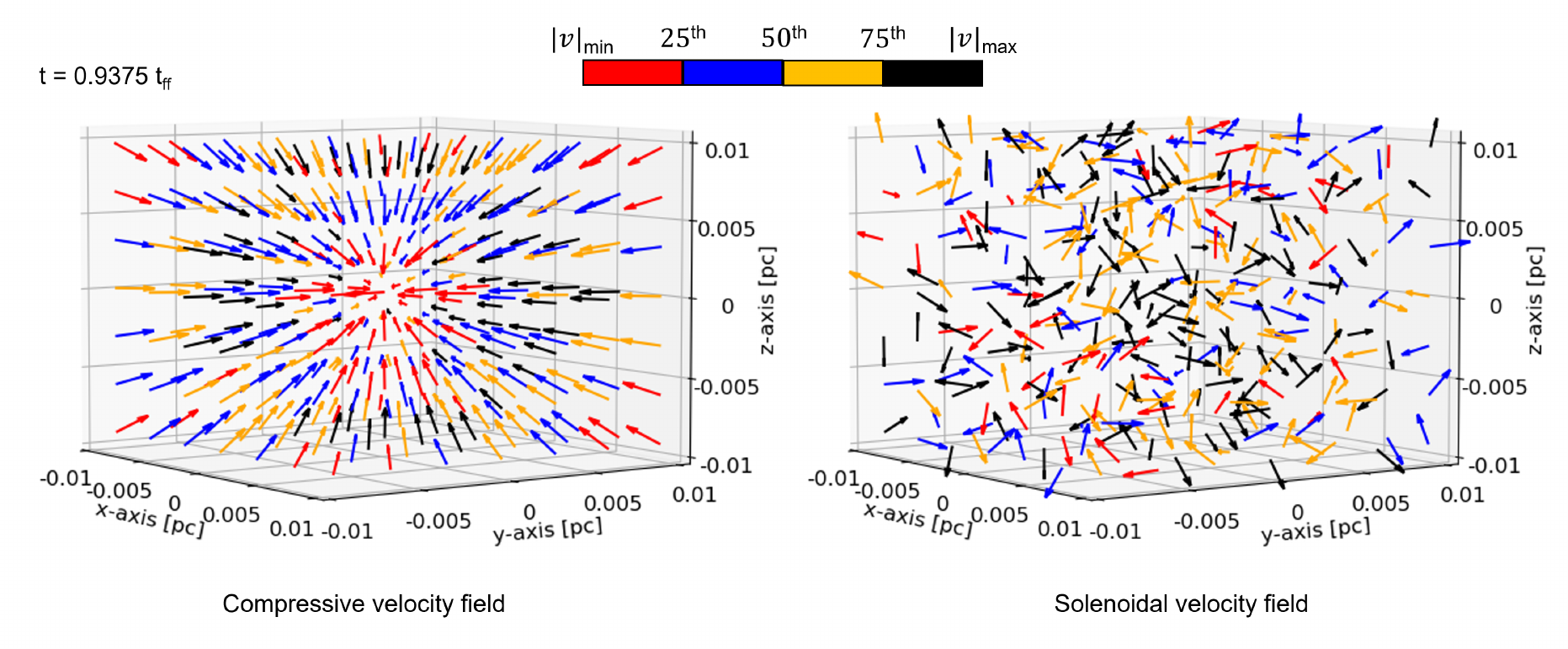}
    \caption{Illustration of the Helmholtz decomposition of total velocity into compressive (left) and solenoidal (right) components, at $t = 0.9375\, \tff$. The compressive component is dominated by infall motions, with the velocity vectors pointing mainly toward the center. The solenoidal component contains a residual contribution from the initial conditions as well as a component driven by the collapse. For clarity, all arrows have the same length, with their colors representing the velocity amplitude in different ranges of percentile, as indicated by the color bar. The percentile is calculated for each velocity field respectively.}
    \label{fig:hel}
\end{figure*}

The two potentials can be calculated from the Green function for the Laplacian:
\begin{equation}
\label{eq.green}
    \begin{aligned}
        \phi(\pmb{r})&=\frac{1}{4\pi}\int\frac{\nabla'\cdot \pmb{v}(\pmb{r}')}{|\pmb{r}-\pmb{r}'|}d^3\pmb{r}',\\
        \pmb{\Phi}(\pmb{r})&=\frac{1}{4\pi}\int\frac{\nabla'\times \pmb{v}(\pmb{r}')}{|\pmb{r}-\pmb{r}'|}d^3\pmb{r}',\\
    \end{aligned}
\end{equation}
where $\pmb{r}$ is the position vector and $\nabla'$ is the nabla operator with respect to $\pmb{r}'$. 
Thus, the decomposition can be rewritten as:
\begin{equation}
        \pmb{v}=-\frac{1}{4\pi}\nabla\int\frac{\nabla'\cdot \pmb{v}(\pmb{r'})}{|\pmb{r}-\pmb{r'}|}d^3\pmb{r}'+\frac{1}{4\pi}\nabla\times\int\frac{\nabla'\times \pmb{v}(\pmb{r'})}{|\pmb{r}-\pmb{r'}|}d^3\pmb{r}',
\end{equation}
Note that Eq.~\eqref{eq.green} basically is a convolution with  Green's function $\left(\frac{1}{4\pi|\pmb{r}-\pmb{r}'|}\right)$. It is convenient to solve Eq.~\eqref{eq.green} in Fourier space, and we do that in this work. The Fourier components of the potential fields are then transformed back to real space to obtain the two velocity components. We illustrate the result of the Helmholtz decomposition of the velocity field for one snapshot of the simulation, corresponding to $t = 0.9375 \tff$ in Fig.\ \ref{fig:hel}.

\subsection{Radial profiles} \label{sec:profile}

The radial profiles of magnetic field, velocity, and density are calculated in two different ways. The first one is {\it shell-averaging}, in which we compute the RMS values of the variables over spherical shells of thickness $\sim 1$ grid cell, centered in the box's center. This allows visualization of how the variables vary as a function of radius. The second is {\it volume averaging}, which computes the average over the full spherical volume out to the indicated radius. Shell averaging is used in Figs.~\ref{fig:vshell}, \ref{fig:Bshell}, \ref{fig:vcvs}, and \ref{fig:ratio}, while we use volume averaging in Figs.~\ref{fig:rho}, \ref{fig:rhoB}, and \ref{fig:m2b} for the purpose of calculating mass-to-flux ratio. We use the subscripts ``shell'' and ``sph'' to distinguish the two cases.

\section{Results} \label{sec:results}

In what follows, we discuss various aspects of the collapse in the MHD simulation, and only occasionally refer to the HD simulation, when comparison to the nonmagnetic case is needed. When no specific reference is made, the MHD case should be assumed. Also, in order to quantify the contraction, we define ``the core'' as the region within the radius at which the spherically-averaged (see Sec.\ \ref{sec:profile}) infall speed becomes maximum. In this region, the density field is roughly flat in the absence of turbulence \cite[e.g.][]{WS85, Naranjo+15}.

\subsection{Anisotropy of gravitational contraction} \label{sec:anisotropy}

\begin{figure*}
\centering
    \includegraphics[width=1.0\linewidth]{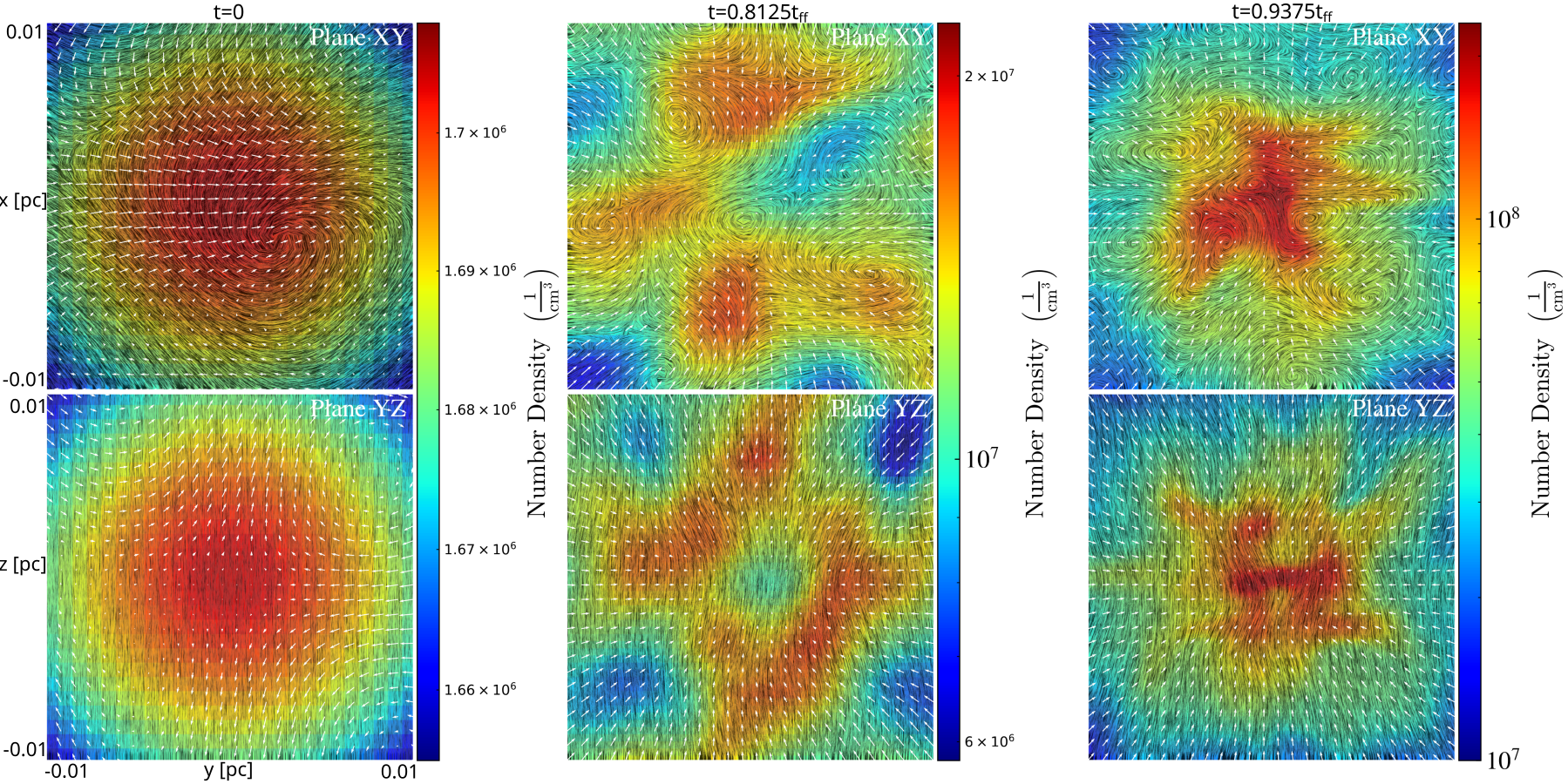}
    \includegraphics[width=1.0\linewidth]{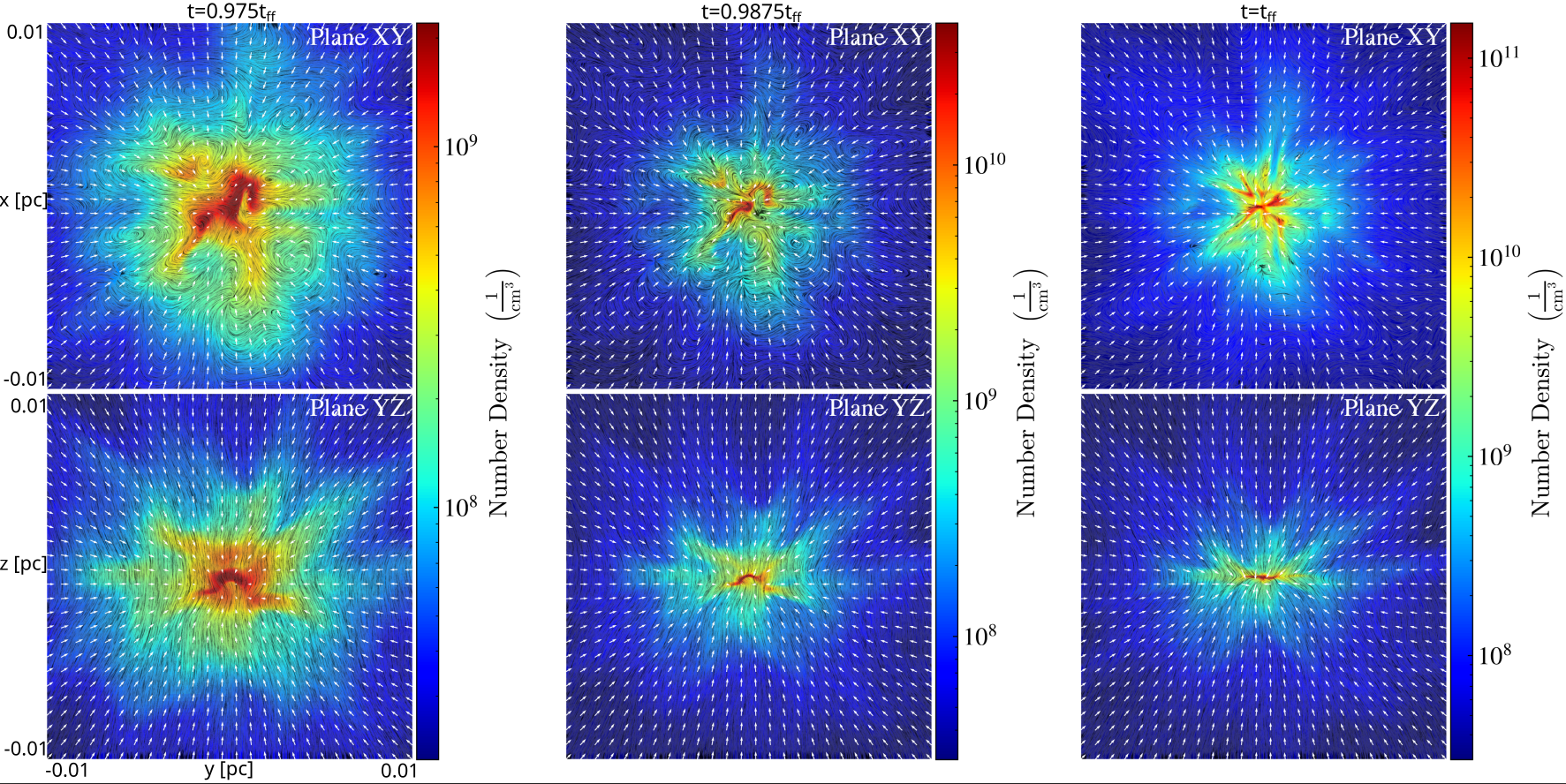}
    \caption{Cross sections through the numerical box along the central $x$--$y$ plane (first and third rows) and along the central $y$--$z$ plane (second and fourth rows) at the indicated times during the gravitational collapse. The color scale represents density. The black lines indicate magnetic streamlines generated from the components of the magnetic field in the corresponding plane, while the white vectors represent the component of the velocity on these planes. Note that the velocity vectors all have the same length, and only depict the direction of the velocity. Note the changing range of the color bar as time proceeds. Furthermore, note that the density field appears less centrally concentrated at $t=0.8125 \tff$ than at $t=0$. This is because the initial condition was a centrally-peaked Gaussian profile. However, at $t=0.8125 \tff$ a Plummer-like profile has been self-consistently established by the collapse, but the size of its central flat-density part is still of comparable size to that of the region shown in the image (cf.\ Fig.\ \ref{fig:rho}), and therefore appears less concentrated than at time $t=0$. At times $t \ge 0.9375 \tff$, the central flat part of the Plummer-like profile is already smaller than the region shown, and thus a central peak is noticeable again in the images.}
    \label{fig:1}
\end{figure*}

Figure \ref{fig:1} shows 2D cross sections of the MHD numerical box over the $z=0$ plane (first and third rows) and over the $x=0$ plane (second and fourth rows; recall the initial magnetic field is parallel to the $z$ axis) at different stages of the collapse, from $t = 0$, i.e., when the collapse begins, to $t = \tff$. The collapse achieves contraction ratios $1, 1/2, 1/4, 1/8, 1/16$, and $1/32$ at $t=0$, $0.8125t_{\rm ff}$, $0.9375t_{\rm ff}$, $0.9750t_{\rm ff}$, $0.9875t_{\rm ff}$, and $t_{\rm ff}$, respectively. Starting from $t=0.8125 t_\text{ff}$, the panels in Fig.\ \ref{fig:1} show the central collapsing region, of size $L_0/2$, where $L_0$ is the initial box size at $t=0$.

At the onset of the collapse ($t=0$), the density structure is nearly isotropic, although, in the $x$-$y$ plane, i.e., the plane perpendicular to the mean magnetic field, the magnetic field (shown by the black lines) is tangled by turbulent motions. As the collapse proceeds, the turbulent eddies (i.e., the vortex-like structures seen in turbulent magnetic fields) undergo compression and their sizes decrease with time. On the other hand, in the $y$-$z$ plane, the density structure is seen to become anisotropic, becoming shorter along the direction of the $z$ axis, causing the formation of a sheet parallel to the $x$-$y$ plane. The collapse in the presence of a mean magnetic field naturally generates an anisotropy \citep[e.g.,] [] {Shu+87}, which provides a magnetic force primarily in the direction perpendicular to the mean field. At later stages, the density structure has contracted strongly, and we observe the turbulent perturbation of magnetic fields, but the anisotropy of magnetic field configuration remains, and an hourglass morphology gradually forms in the central region.

\begin{figure*}
\centering
	\includegraphics[width=1.0\linewidth]{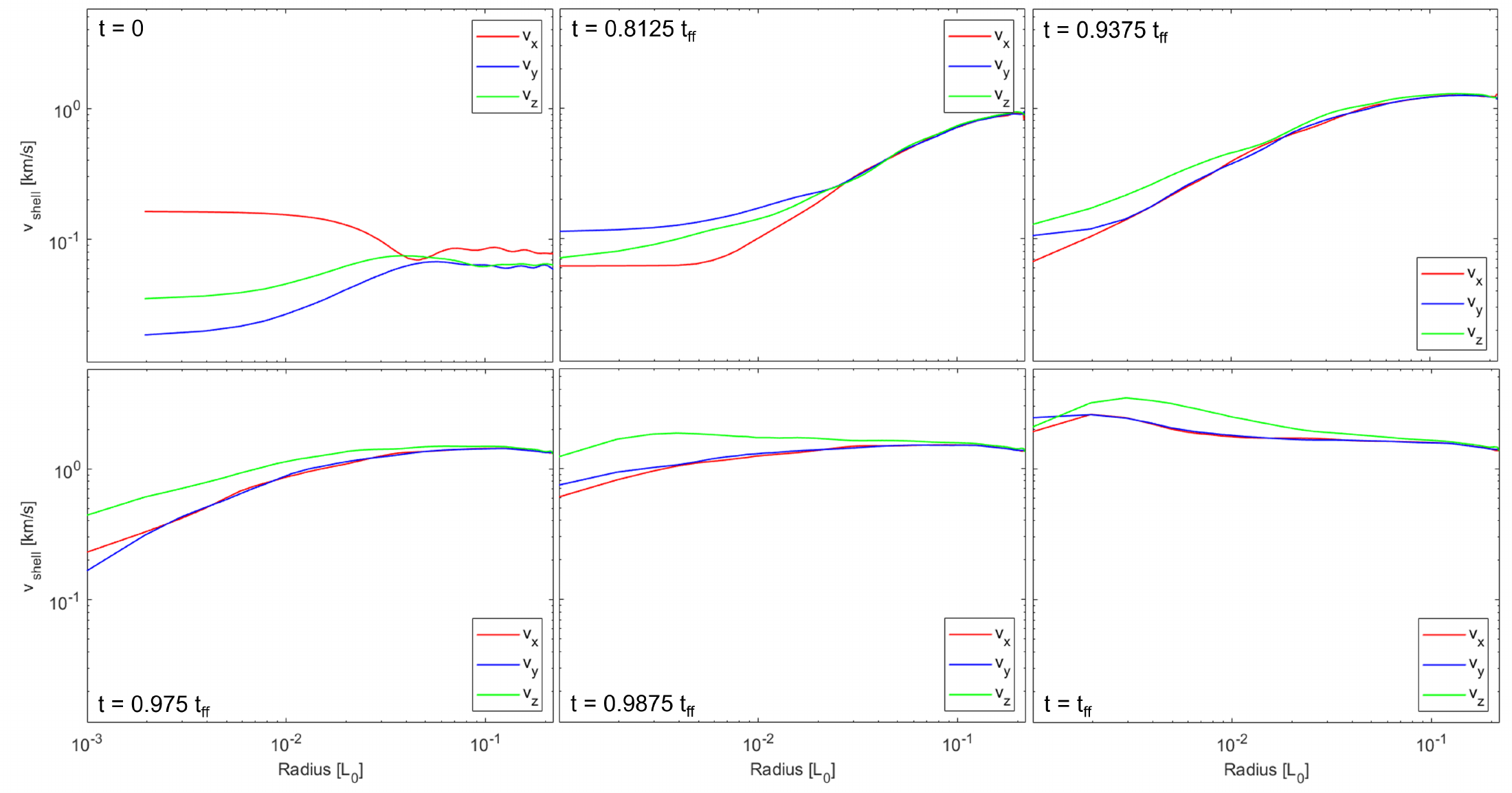}
    \caption{Radial profiles of the shell-averaged RMS values of the three components of the velocity field $v_x$, $v_y$, and $v_z$, at the indicated timed during the collapse, in units of the free-fall time corresponding to the mean density of the central Jeans mass at $t = 0$. The radius is normalized by the box length, $L_0 = 0.1$ pc.}
    \label{fig:vshell}
\end{figure*}

Figure \ref{fig:vshell} presents the RMS values of the $x$, $y$, and $z$ components of flow velocity, $v_x$, $v_y$, and $v_z$, as a function of $R/L_0$, where $R$ is the radius of a spherical region centered at the center of our simulation box. Taking advantage of the approximate spherical symmetry, the RMS velocity value is averaged over spherical shells of thickness $\approx1$ grid cell, which is denoted as $v_{\rm shell}$. Initially (at $t=0$), $v_x$ exhibits a larger amplitude in the central region ($R\lesssim 0.4L_0$), while $v_y$ and $v_z$ show the opposite trend. This is just the manifestation of the random initial turbulent velocity field. At later times ($t/\tff \ge 0.9375$), however, $v_z$ is seen to be somewhat larger than the other two components, indicating the unrestricted collapse in this direction. Along the $x$ and $y$ directions, instead, the collapse is slightly delayed by the magnetic support.

At not-too-advanced stages of the collapse ($t \lesssim 0.9750t_{\rm ff}$), we see that, under the effect of gravity, the contraction is faster in the outer region and decays towards the inner region.
This is consistent with the well-known solution of nonmagnetic spherical prestellar collapse, for which the central parts of the sphere (the region where the density is nearly flat), the infall speed scales linearly with radius, while in the outer regions, where the density drops off as $r^{-2}$, the infall speed becomes constant with radius \citep[e.g.,] [] {WS85}. According to this solution, the inflection radius at which the speed changes from linear to constant with radius approaches the center, reaching the latter at the moment of formation of the singularity (the protostar). 

\begin{figure*}
\centering
	\includegraphics[width=1.0\linewidth]{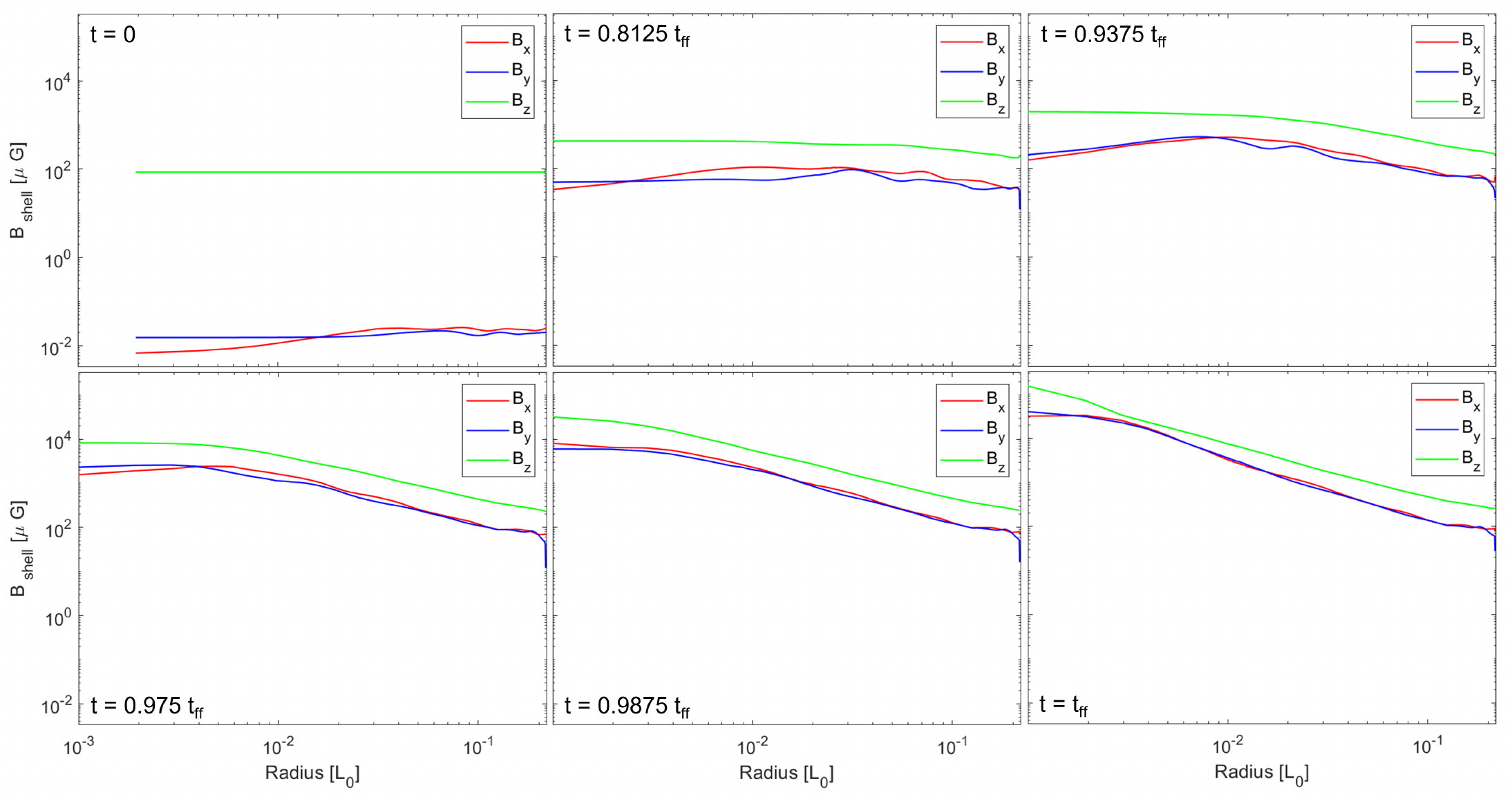}
    \caption{Similar to Fig.\ \ref{fig:vshell}, but for the radial profiles of the shell-averaged RMS values of the three components of the magnetic field $B_x$, $B_y$, and $B_z$.}
    \label{fig:Bshell}
\end{figure*}

The radial variation of the magnetic field strength is presented in Fig.~\ref{fig:Bshell}. Due to the presence of an initial large-scale magnetic field along the $z$-direction, the anisotropy in magnetic field distribution persists through the collapse (see Fig. \ref{fig:Bshell}), with $|B_z| > |B_x|\approx |B_y|$.
 However, unlike the velocity field, the magnetic field is always stronger in the inner region than in the outskirts during the collapse process. Compared to the initial conditions, the mean magnetic field is amplified via compression by two orders of magnitude in the inner region, while in the outskirts it stays nearly unchanged.

\subsection{Turbulence amplification by the collapse} \label{sec:grav_amp}

As discussed in Sec.\ \ref{ssec:Hel}, and illustrated in Fig.~\ref{fig:hel}, we use a Helmholtz decomposition to separate the velocity field in its solenoidal and compressive components. 
The compressive velocity is seen to be oriented toward the collapsing center as it is dominated by the infall motions, but we note that it can also contain turbulent (non-infall) motions. The solenoidal component, on the other hand, is randomly oriented, as it corresponds to purely turbulent motions. As the externally driven initial turbulence decays over roughly one crossing time, the solenoidal turbulence seen at later times must be generated by gravitational contraction.

\begin{figure*}
\centering
	\includegraphics[width=1.0\linewidth]{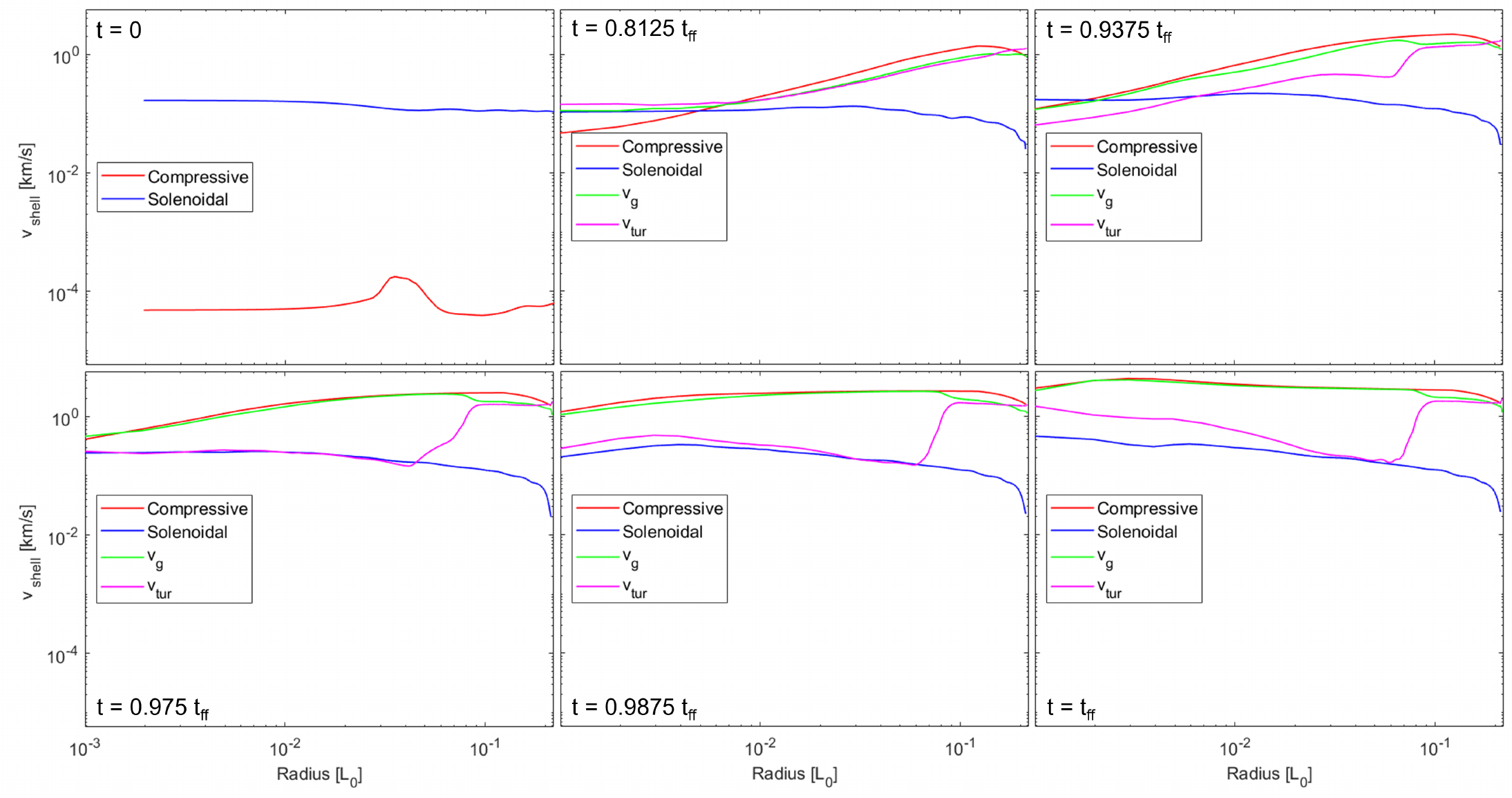}
    \caption{Similar to Fig.\ \ref{fig:vshell}, but for the radial profiles of the shell-averaged RMS values of the solenoidal, compressible, infall, and turbulent velocity fields.}
    \label{fig:vcvs}
\end{figure*}

The radial profiles of the RMS values of the solenoidal ($v_{\rm s}$) and compressive ($v_{\rm c}$) velocity components, averaged over spherical shells, are presented in Fig.~\ref{fig:vcvs}. At $t=0$, the large solenoidal component corresponds to the initial solenoidal imposed turbulence, while the compressive component is virtually nonexistent, since the collapse motions have not started yet. However, at the more advanced times shown in the figure, the compressive component becomes dominant. The maintenance and moderate increase of the solenoidal velocity at these later stages of the collapse is accounted for by the driving by gravity. However, it can be seen that the increase in the solenoidal component is very mild. In fact, this component remains almost at the same level throughout all snapshots shown.

\begin{figure*}
	\includegraphics[width=1.0\linewidth]{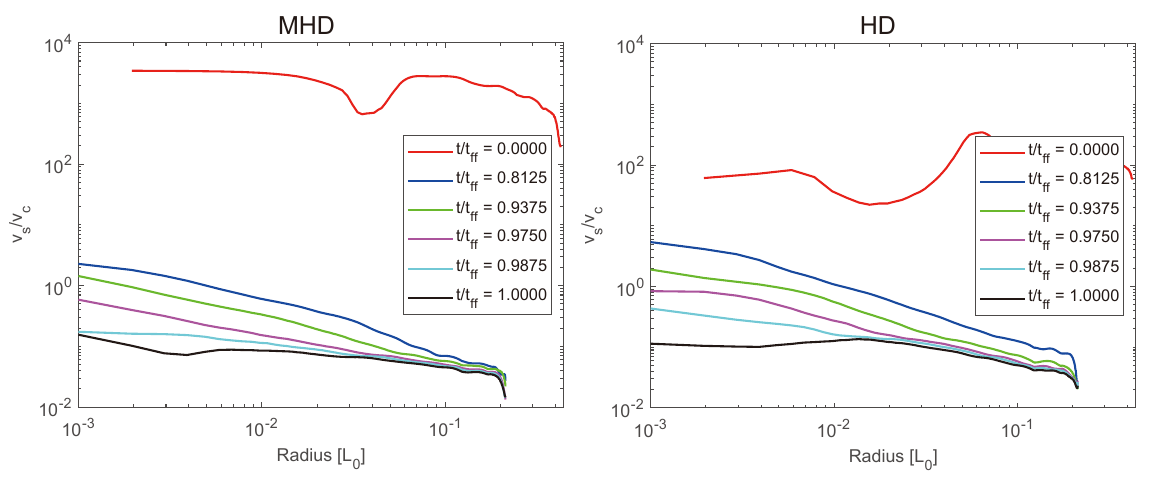}
    \caption{Radial profile of the shell-averaged ratio of the RMS solenoidal to compressive components for the MHD (left) and hydrodynamic simulations (right).}
    \label{fig:ratio}
\end{figure*}

The ratio of the solenoidal to compressive velocity components in the MHD run is presented in the left panel of Fig. \ref{fig:ratio}. A higher ratio is seen at smaller $R$ due to the slower contraction speed in the inner region, as shown in Fig. \ref{fig:vcvs}. When $t\ge 0.9375 t_{\rm ff}$, the ratio becomes less than unity throughout the radial range shown, indicating that the compressive component becomes dominant over the solenoidal component almost everywhere in the core.

With our interpretation of the solenoidal modes as the turbulent component of the total motions, this low amplification level of the solenoidal component would imply that the trend towards equalization of the eddy turnover rate and the collapse rate predicted by \citet{RobG12}, or the significant amplification of initially subsonic turbulence by gravitational compression observed in various other studies  \citep[e.g.,] [] {Sur12, Higa21, Henn21}, 
is not realized in our simulations. Neither is the pseudo-virial state observed by \citet{GV20}, in which the kinetic energy in the infall (compressive) motions was roughly twice the kinetic energy in the turbulent motions. Those authors attributed this regime to the increasing infall kinetic energy, which constituted an increasing driving rate for the turbulent motions. Instead, in the present simulations, the solenoidal motions seem to remain almost constant in time, and uniform in radius.

At face value, this result would suggest an inefficient amplification of turbulence in our simulations. This could be due to insufficient resolution. However, in the Appendix we show that our results do not vary strongly when increasing the resolution by two additional refinement levels, suggesting that our results are converged.




Another possibility is that the magnetic tension may also suppress the amplification of turbulence. Mediated by the large-scale ordered magnetic field threading the collapsing region, the angular momentum of the compressed turbulent eddies may be transferred along the magnetic field away from the collapsing region, suppressing the gravity-driven solenoidal motions there.

To examine the magnetic effect on the amplification of turbulence,
we make the comparison with the HD simulation in Fig.~\ref{fig:ratio}.
We find that in the HD case, the solenoidal fraction is generally larger than in the MHD case throughout the radial range, and at all times $t>0$, although only mildly, thus not constituting an important effect on the generation of solenoidal motions by the collapse. On the other hand, at $t=0$, we see that the solenoidal fraction is significantly lower in the HD case, suggesting that the main role of the magnetic field at transonic Mach numbers is to prevent the transfer of energy from the solenoidal to the compressible modes during the pre-gravity driving stage of our simulation. This result is consistent with the early finding by \citet{VS+96}, that the maintenance of solenoidal energy requires the presence of vorticity-generating forces such as the Lorentz or Coriolis forces. Nevertheless, the increase of compressible energy in the HD case is not enough to make a significant difference at later times.

The above tests suggest that the low amplification level of the solenoidal component is not due to insufficient resolution nor to the presence of the magnetic field. A third possibility is that a significant fraction of the turbulent energy generated by the collapse with our setup is {\it not} in the solenoidal modes, but rather in the compressible ones. Indeed, a measurement of the turbulent energy fraction using the same method as in \citet{GV20} (Guerrero-Gamboa \& V\'azquez-Semandeni, in prep.) shows that, although the turbulent energy fraction is indeed somewhat smaller in the magnetic case, the difference with the hydrodynamic case is much smaller than our measurements here would suggest. Since that method takes into account both the solenoidal and compressible energies in the estimation of the turbulent fraction, our results here suggest that an important fraction of the kinetic energy generated by the collapse is in compressible, rather than solenoidal modes.

To test for this, we define the {\it infall} (or {\it gravitational}) velocity as the component of the velocity vector along the direction of the gravitational acceleration vector: 
\begin{equation}
\vgvec \equiv {\bm v} \cdot \hat{{\bm g}},
\label{eq:vg}
\end{equation}
where ${\bm v}$ is the total velocity vector, ${\bm g} \equiv \nabla \phi$ is the gravitational acceleration vector, $\phi$ is the gravitational potential, and $\hat{\bm g}$ is the unit vector along ${\bm g}$. Making the approximation that $\vg$ is the velocity driven by the gravitational acceleration, we then define the turbulent component as
\begin{equation}
\vturbvec \equiv {\bm v} - \vgvec.
\label{eq:vturb}
\end{equation}
Note that this is only an approximation, as both the truly turbulent velocity and the solenoidal velocity $\vsvec$ are not necessarily perpendicular to the gravitational acceleration vector $\bm g$. Therefore, the approximation given by Eq.\ \eqref{eq:vturb} may underestimate the true turbulent velocity. 

The shell-averaged RMS values of these two velocities (denoted $\vg$ and $\vturb$) are also shown in the frames corresponding to $t \ge 0.8125 \tff$ of Fig.\ \ref{fig:vcvs}, and can be compared to the RMS values of the compressible and solenoidal components, which we respectively denote by $\vc$ and $\vs$. From that figure, we see that, at late times and/or large radii (i.e., when the infall speed is large), $\vc \gtrsim \vg$ in general, implying that there are compressive motions that do not correspond to infall. Also for late times/large radii (although not exactly in the same radial ranges as before), $\vturb > \vs$, indicating that there is a non-solenoidal (therefore compressible) contribution to the turbulent velocity. Both of these results imply that there is a significant compressible contribution to the turbulent speed.

However, we also note that, at early times (the frames for $t = 0.8125 \tff$ and $t = 0.9325 \tff$), there are regions in the inner parts of the core where $\vc < \vg$ and/or $\vturb < \vs$. This happens because of the limitation stated above, that our definition of $\vgvec$ is contaminated by a contribution of the turbulent velocity vector in the direction of the gravitational acceleration. Therefore, at times or radial ranges where the infall speed is small, this contamination dominates $\vgvec$. However, once the infall speed is dominant, the presence of a significant compressible component in the turbulent velocity is clear.

\subsection{$B-\rho$ correlation} \label{sec:B-n}

The correlation between the magnetic field strength and the density is an important aspect of the theory of star formation. However, many different effects contribute to the scaling between the magnetic field strength and the density upon compressions, of either turbulent or gravitational origin. 

Under the ideal MHD condition, and in the presence of a weak field, a spherical core contracts isotropically, maintaining its spherical geometry. If both the mass, $M \propto\rho R^3$, and the magnetic flux, $\Phi \propto B R^2$, are conserved, the field is expected to scale as $B\propto \rho^{2/3}$ \citep[e.g.] [, Ch. 24] {Shu92}. On the other hand, when the 
 the field is strong, the initial contraction mainly takes place along the mean magnetic field, and the gas settles into a flattened cylindrical structure perpendicular to the mean field. In the limit of very strong fields, this produces an increase of the density at constant field strength \citep[e.g.,] [] {Mestel65, Hartmann+01, VS+11}. However, if accretion along field lines continues to increase the gas mass responsible for generating the gravitational potential \citep{Hartmann+01, VS+11}, then the flattened cloud must contract radially to some extent, producing the well-known hourglass shape. If the thickness is assumed to be determined by the hydrostatic balance between thermal pressure and gravity in the direction of the mean field, then $B \sim \rho^{1/2}$ \citep{Mousch76, Scott_Black80, Crut99}. 

However, $B\propto \rho^{1/2}$ can also be the result of ambipolar diffusion \citep{Mousch76, Ciolek_Mousch94}, reconnection \citep{Sant10,LEC12,Xul20a,Xul20b}  or other forms of diffusion of the magnetic field, which cause the breakdown of flux freezing and partial decorrelation between $B$ and $\rho$. 

 On the other hand, in the purely turbulent case without self-gravity, \citet[] [hereafter PV03] {Passot_VS03} considered the scaling of the magnetic pressure ($\sim B^2$) with density for the various modes of ``simple'' (nonlinear) MHD waves, showing that each mode produces a different scaling, without the need to invoke any form of diffusion or dissipation. Specifically, they showed that, when the slow mode dominates, a scaling of the form $B^2 = c_1 - c_2 \rho$ emerges, where $c_1$ and $c_2$ are positive constants. However, the slow mode disappears at large density, when $\rho > c_1/c_2$. When the fast mode dominates, those authors showed that a scaling of the form $B^2 \propto \rho^2$ arises. Finally, they showed that the pressure for a circularly polarized Alfv\'en wave is of the form $B^2 \propto \rho^{\gamef}$, with $\gamef \approx 2$ at large Alfv\'enic Mach number $\Ma$, $\gamef \approx 3/2$ at moderate $\Ma$, and $\gamef \approx 1/2$ at low $\Ma$. From all these different scalings, PV03 concluded that the $B$-$\rho$ correlation is not unique, and depends on the {\it history} of wave passages through a given location in the flow, rather than simply on the local value of the density. Nevertheless, the generic form of the $B$-$\rho$ scaling is to be flat at low densities and to increase as some power of $\rho$ in the range $1/2 < \gamef <1$ at high density, in qualitative agreement with observations \citep[e.g.,] [] {Crutcher12}.

\begin{figure*}
	\includegraphics[width=1.0\linewidth]{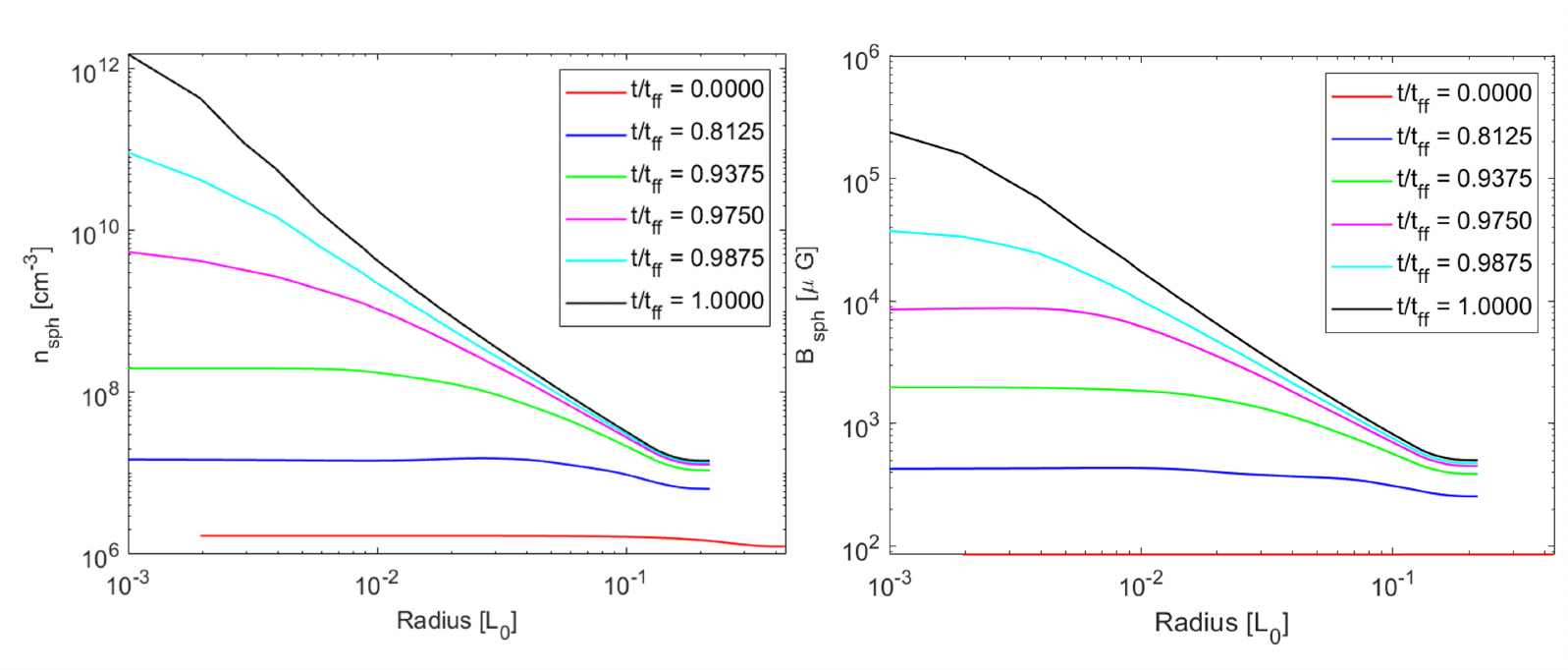}
    \caption{Evolution of the radial profiles of volume-averaged $\rho_\text{sph}$ and $B_\text{sph}$.}
    \label{fig:rho}
\end{figure*}

 With all the above background, we can now proceed to discuss the $B$-$\rho$ correlation in our simulation of magnetized turbulent gravitational collapse. In Fig.~\ref{fig:rho}, we compare the radial profiles of volume-averaged density 
$\rho_\text{sph}$ and magnetic field strength $B_\text{sph}$ within the sphere of radius $R$ (normalized by $L_0$). As a result of gravitational collapse, the density and the magnetic field lines are significantly compressed in the central region. The density can be compressed up to six orders of magnitude, while the magnetic field is moderately amplified by three orders of magnitude approximately. With a similar distribution of $\rho_\text{sph}$ and $B_\text{sph}$ at different stages of the gravitational collapse, the amplification of magnetic field strength is mainly attributed to the compression generated by the contraction.

\begin{figure*}
	\includegraphics[width=1.03\linewidth]{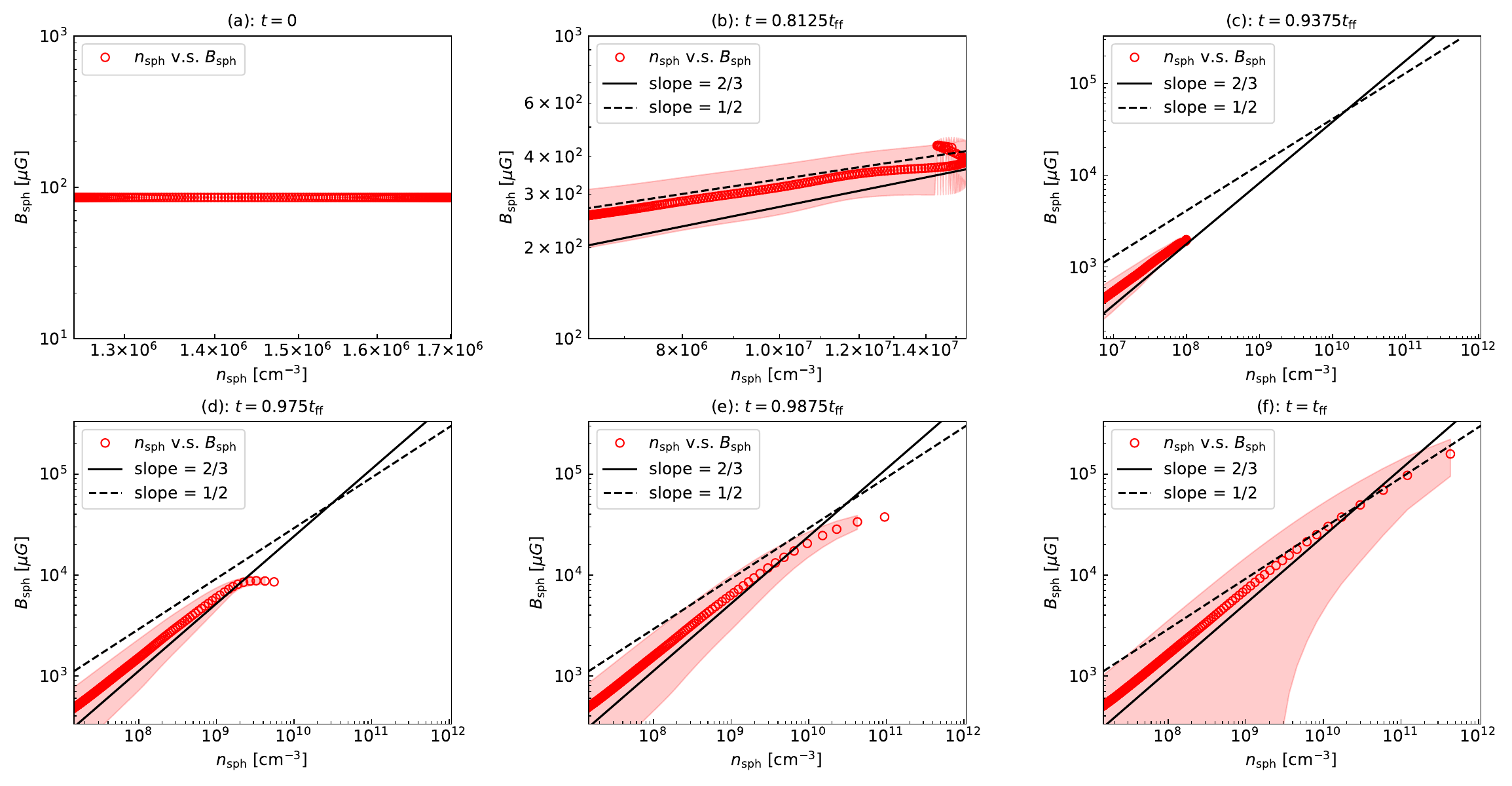}
    \caption{Evolution of the correlation between the spherically averaged gas density and the magnetic field strength. The shaded area gives the standard deviation of magnetic field strength averaged over the corresponding spherical volume. The kink in the curve at $t = 0.8125 \tff$ is caused by the fact that, due to the turbulent motions, the collapse is not precisely focused at the center, and so neither the maximum density nor the maximum field strength occur inside the innermost spheres used to generate these plots. At later times this is not noticeable because the collapse is so advanced that the offset from the center is smaller than the size of the innermost sphere.}
    \label{fig:rhoB}
\end{figure*}

The correlation between $B_\text{sph}$ and $\rho_\text{sph}$
is presented in Fig. \ref{fig:rhoB}. At $t=0.8125 t_\text{ff}$, we find $B\propto \rho^{1/2}$, presumably as a result of the anisotropic contraction within the entire collapsing region (see Fig. \ref{fig:1}). 
However, already at the panel corresponding to $t = 0.9375t_\text{ff}$, the slope of the correlation is very close to $2/3$, although, starting from $t=0.975 t_\text{ff}$, we see a kink in the slope from $B\propto \rho^{2/3}$ to a shallower value at the highest densities, beyond which the slope approaches $B\propto \rho^{1/2}$. 

The flattening of the $B-\rho$ scaling at the highest densities (i.e., toward the central region) can be due either to the mode of collapse or to the enhancement of diffusion. As mentioned above, in the spherically collapsing, ideal (nondiffusive) case, the expected value of the exponent is $\sim 2/3$, while, if it occurs first onto a sheet-like cloud whose thickness is determined by hydrostatic equilibrium, the expected exponent is $\sim 1/2$. On the other hand, in the diffusive case with spherical collapse, a flattening of the slope from $\sim 2/3$ to $\lesssim 1/2$ is expected when diffusion becomes important. If the diffusion is enhanced in the inner regions because of the larger infall speeds (and consequently, a larger injection rate) there, then one would expect the slope to be near 1/2 in the central parts.

As seen in Fig.\ \ref{fig:rhoB}, the logarithmic slope of the $B$--$\rho$ curve in our simulation is $\sim 1/2$ at early times (see the panel corresponding to $t =  0.8125 \tff$), but then transitions to $\sim 2/3$ at later times and low densities, suggesting a transition from a mostly planar collapse at early times to a roughly spherical one at late times. This is confirmed by the morphology of the density, velocity, and magnetic fields seen in the frame corresponding to the $y$-$z$ plane $t = 0.8125 \tff$ in Fig.\ \ref{fig:1}. In this panel, a horizontal, flattened, intermediate-density sheet is observed to have formed, and to be contracting radially, as indicated by the velocity arrows. This is precisely the configuration for which a slope $\sim 1/2$ is expected. Instead, at later times, the central layer has disappeared, and the collapse appears to be roughly isotropic, consistent with the slope of $2/3$ observed for these times. However, at the highest densities ($n \gtrsim 10^{10} \pcc$) at the later times, a slope of $\sim 1/2$ is observed (see the panel corresponding to $t = \tff$), which can be attributed to reconnection diffusion based on the numerical diffusion. We conclude that both the geometry and the reconnection diffusion play a role in the determination of the mean $B$--$\rho$ correlation during the collapse.

\begin{figure*}
	\includegraphics[width=0.49\linewidth]{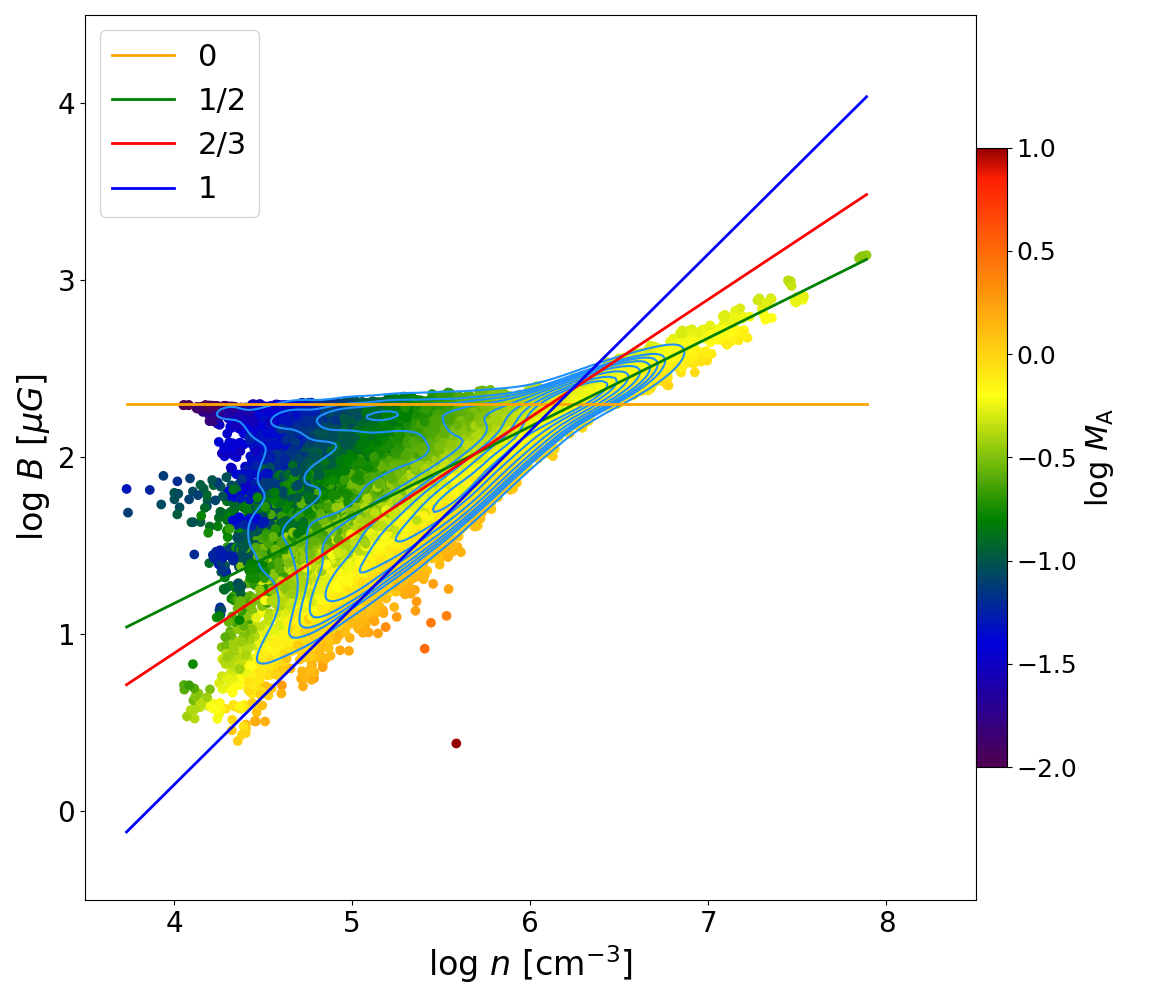}
	\includegraphics[width=0.49\linewidth]{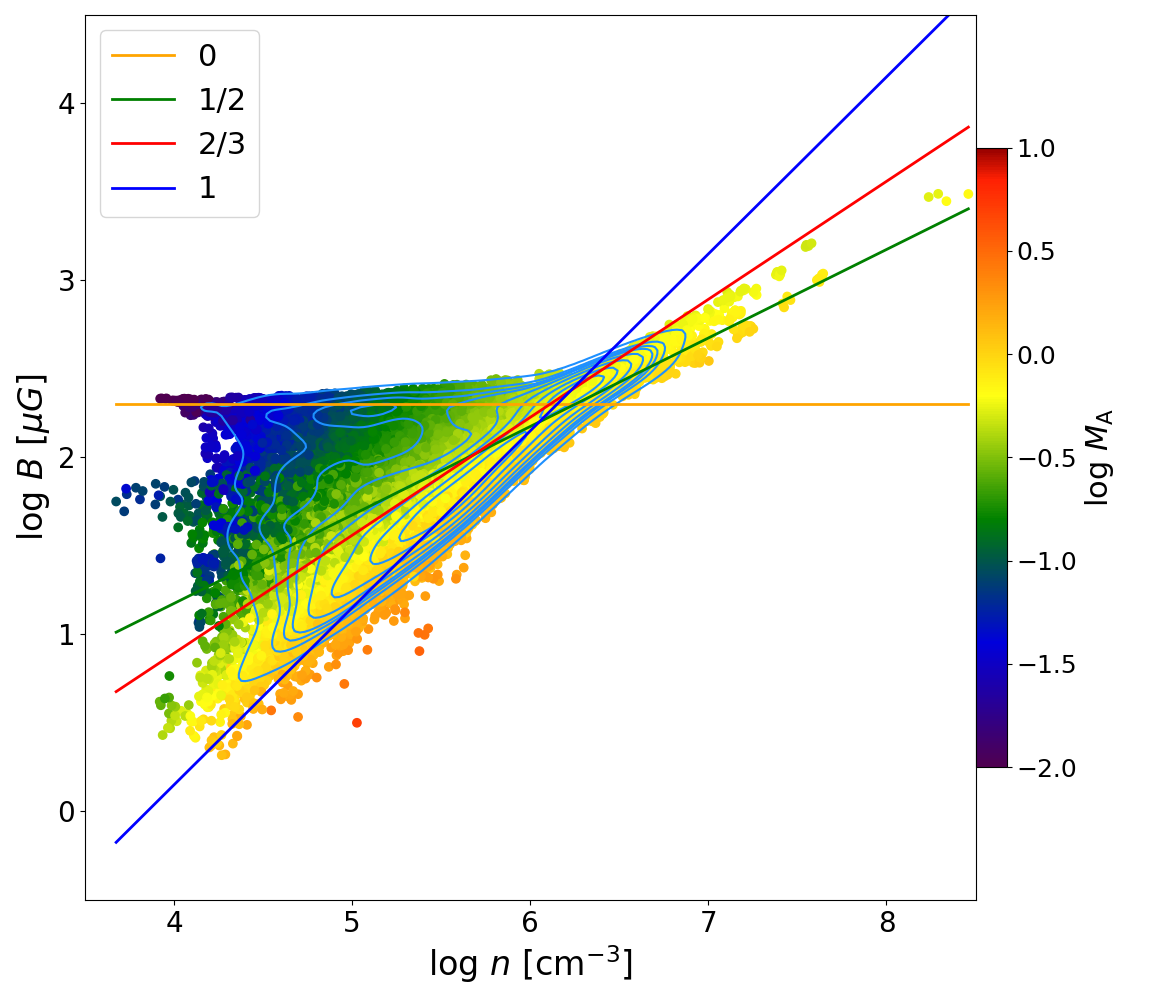}
    \caption{Two-dimensional histogram of simulation points in the $B-n$ space at times $t = 0.9375 t_{\rm ff}$ (left panel) and $t = \tff$ (right panel). The dots are colored by the corresponding value of the Alfv\'enic Mach number, $M_\mathrm{A} = |v|\sqrt{4\pi\rho}/|B|$. The lines represent slopes of 0, 1/2, 2/3, and 1, as indicated in the inset.}
    \label{fig:rhoB_scatter}
\end{figure*}

Finally, in Fig.\ \ref{fig:rhoB_scatter}, we show the magnetic field-density scatter plot, taking each cell of the simulation as a dot in the plot, and the two-dimensional probability density (in contours), for $t = 0.9375 \tff$ (left panel) and $t = \tff$ (right panel). The dots are colored with the value of the Alfv\'enic Mach number, $|v|\sqrt{4 \pi n}/B$ in the corresponding cell, using the magnitude of the velocity. In the contours, we can distinguish various superposed regimes. First, at low density, the contours cover an extended area, but they concentrate around a zero-slope scaling characteristic of the slow mode (indicated by the orange line), and a scaling with a nearly unit slope characteristic of the fast mode (blue line). It is also noteworthy that the zero slope corresponding to the slow mode occurs for large values of $B$, or, equivalently, for low values of $\Ma$, when the inertia of the turbulent motions is low compared to the magnetic forces. Conversely, the unit slope corresponding to the fast mode occurs for small values of $B$, or, equivalently, for large values of $\Ma$, implying that the inertia of the turbulent motions overwhelms the magnetic forces.

Second, at high densities, the scatter is strongly reduced and, at $t = 0.9375 \tff$, a single scaling with slope 1/2 is observed, suggesting that diffusion dominates the scaling. However, at $t = \tff$, a significant fraction of the high-density points is still close to the slope of 2/3, suggesting that, at this stage, the collapse is so fast that diffusion cannot completely erase the signature of spherical collapse, in agreement with the conclusion of \citet{GV20} that, as the collapse accelerates, the turbulent cascading rate and the dissipation cannot ``catch up'' with the energy injection, thus preventing the equipartition levels of turbulence suggested by \citet{RobG12}.


\subsection{Mass-to-flux ratio}

The mass-to-flux ratio, $M/\Phi$, is considered an important diagnostic of the energy balance within a core, determining whether it can be supported by the magnetic field against its self-gravity \citep{Mestel_Spitzer56}. Here we compute it at radius $R$, normalized to the critical value for cylindrical geometry, $(M/\Phi)_\text{crit} = 2 \pi \sqrt{G}$ \citep{Nakano_Nakamura78}, as:
\begin{equation}
\lambda_c \equiv \frac{M/\Phi}{(M/\Phi)_\text{crit}}=
    2\pi\sqrt{G}~\frac{\oint_{r<R}\rho(x,y,z)d\pmb{x}^3}{\pi R^2 \bar{B}} 
\end{equation}
where $G$ is the gravitational constant, $r$ defines the distance from each grid cell to the center of the simulation box, and $\bar{B}$ is the magnetic field strength averaged over a sphere with radius $R$. When $\lambda_c>1$, the core is magnetically {\it supercritical} (i.e., the gravitational potential energy exceeds the magnetic energy), while if $\lambda_c<1$, the core is subcritical (i.e., its gravitational potential energy is smaller than the magnetic energy). $M/\Phi$ of the entire box is approximately conserved. 

However, as first pointed out by \citet[] [hereafter VS+05] {VS+05}, and more recently quantified by \citet{Gomez+21}, under ideal MHD conditions, the subregions or fragments of a clump or core will in general have a lower value of $\lambda_c$ than the whole clump. VS+05 arrived at this conclusion by considering the limiting cases that bracket the condition of any fragment of a clump. On one extreme, if the {\it whole} clump contracts to a smaller radius under ideal MHD conditions (i.e., with no diffusion of the magnetic field), then the mass-to-flux ratio is conserved, because of the flux-freezing condition and because the mass is the same. On the other end, if one considers only a subregion of the original clump, with the same density and magnetic field strength as the whole clump, then the mass varies as $R^3$, while the flux varies as $R^2$, and so $M/\Phi$ varies as $R$. Therefore, the mass-to-flux ratio of an arbitrary fragment ($\lambda_{\rm f}$) within the cloud is constrained to lie within these two extremes, and so we can write
\begin{equation}
\lambda_c \left(\frac{R_{\rm f}} {R_c}\right) \le \lambda_{\rm f} \le \lambda_c,
\label{eq:limits_lambda}
\end{equation}
where $\lambda_c$ and $R_c$ are respectively the mass-to-flux ratio and the radius of the whole clump, and $R_{\rm f}$ is the radius of the fragment. In addition, \citet{Gomez+21} analytically showed that, for any centrally concentrated sphere with a ``reasonable'' density profile ($\rho \propto r^{-p}$, with $p<3$) and with the magnetic field scaling with density as $B \propto \rho^\eta$, then the mass-to-flux ratio scales as 
\begin{equation}
\frac{M}{\phi} \propto r^{1-p(1-\eta)}.
\label{eq:M2F_of_r}
\end{equation}
In particular, then, for $\eta >(p-1)/p$, $M/\phi$ decreases inwards, and so there is always a certain inner region that will appear magnetically subcritical even if it is embedded within a supercritical larger region. 

\begin{figure}
	\includegraphics[width=1.0\linewidth]{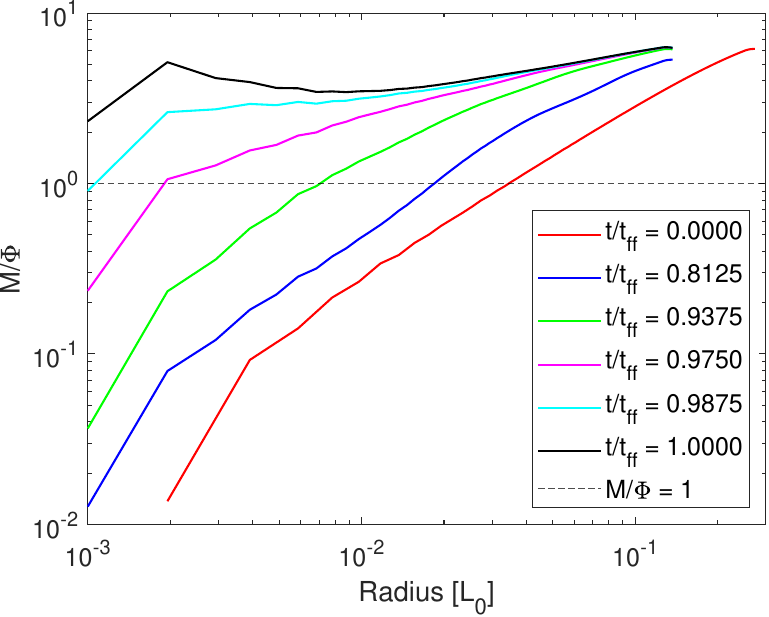}
    \caption{Normalized mass to flux ratio $M/\Phi$ over each sphere centering at the center of the simulation box. $M/\Phi>1$ means supercritical condition, while $M/\Phi<1$ is subcritical.}
    \label{fig:m2b}
\end{figure}

In Fig.\ \ref{fig:m2b} we show the radial profiles of the normalized mass-to-flux ratio out to the indicated radius at various times. In agreement with the theoretical predictions of VS+05 and \citet{Gomez+21}, we see an outwards increase of $\lambda_c$, and a transition from a supercritical outer region to a subcritical inner region for $t \leq 0.975 t_\text{ff}$. Magnetic fields are strongly amplified by compression towards the center (see Fig.\ \ref{fig:Bshell}). 

In addition, the decrease of $M/\Phi$ with decreasing radius is consistent with the observational finding by \citet{Crut09}. A supercritical molecular envelope is also seen in observations (e.g., \citealt{ChiLi22}). Note, however, that the magnetically subcritical inner region shrinks with time as a consequence of the global collapse of the core, which compresses this inner subcritical region.\footnote{It is worth noting, however, that this should not be interpreted as if somehow the pressure (either thermal or magnetic) in the central part were able to counteract the collapse of the entire core. It is not, since the region is increasing its density (i.e., the core {\it is} collapsing). It is simply not collapsing {\it on its own}, but rather just undergoing compression from the infalling outer material. Moreover, since the velocity profile during the prestellar stage is smooth, no shock develops at the boundary of the core's inner region during the prestellar stage. The shock appears simultaneously with the formation of the singularity, at the transition from the pre- to the protostellar stage.} At $t \approx t_\text{ff}$, the entire range of radii seen in Fig.\ \ref{fig:m2b} is supercritical.

\subsection{Effect of the magnetic field on the collapse rate} \label{sec:B_on_coll_rate}

\begin{figure}
	\includegraphics[width=1.0\linewidth]{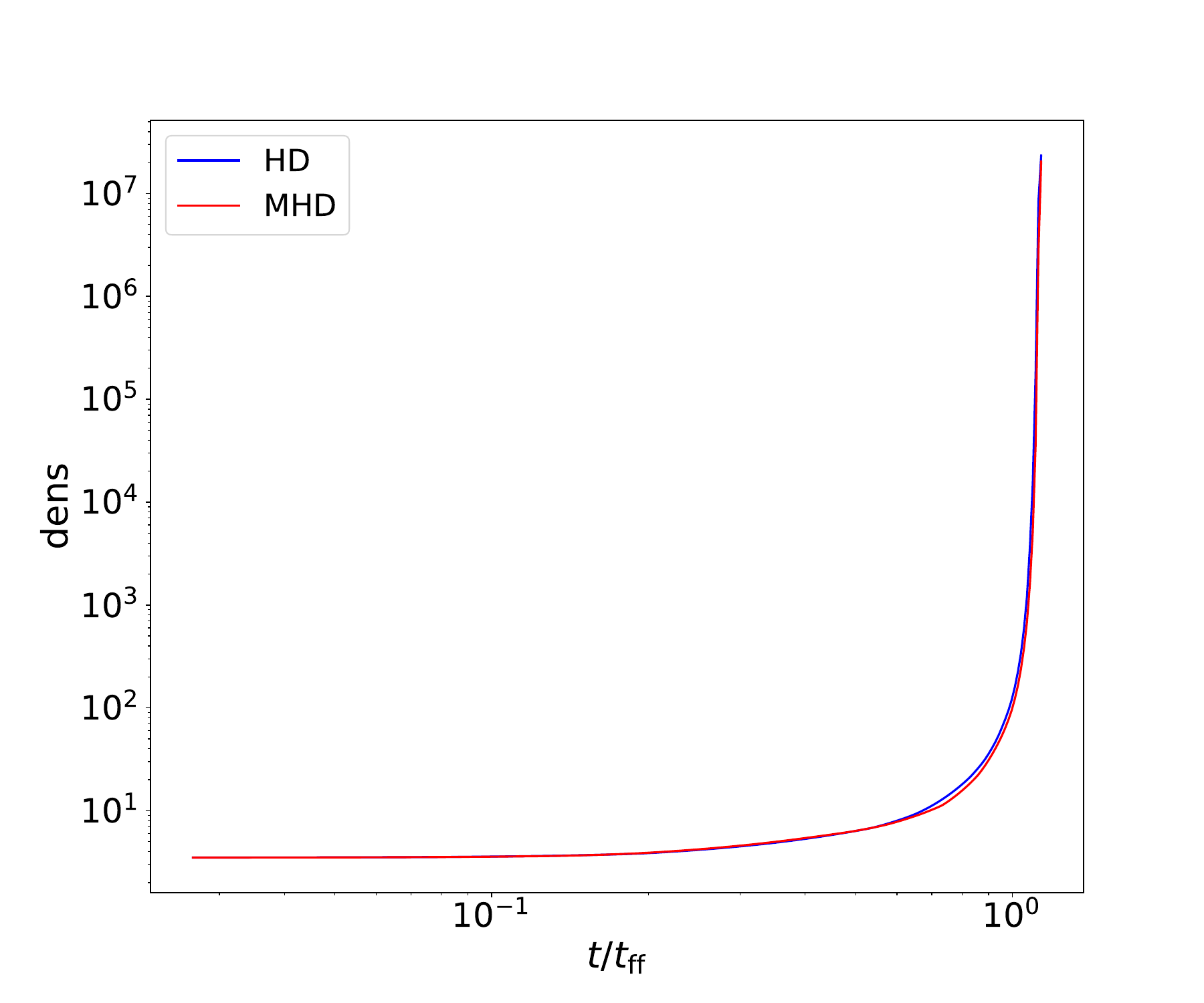}
    \caption{Evolution of the density maximum for the MHD and HD runs. The two runs are seen to behave almost indistinguishably.}
    \label{fig:coll_rate}
\end{figure}

The result from the previous section, that the inner region of the core is magnetically subcritical, may suggest that the collapse might be delayed somewhat by the added magnetic pressure in that region.  To test for this, in Fig.\ \ref{fig:coll_rate} we show the time dependence of the density peak in both the MHD and the HD runs. Somewhat surprisingly, the density peak seems to increase at essentially the same rate in the two simulations, in spite of the existence of the inner subcritical region. This is understandable because the whole numerical box is strongly supercritical, with a total mass-to-flux ratio at the outer edge $\lambda \sim 6$. Therefore, the inner region is being compressed by the infall of the envelope, in spite of being locally subcritical.

\section{Discussion} \label{sec:disc}

\subsection{Radial distribution of the turbulent motions. Is the core ``coherent''?}

Our decomposition of the velocity field in its compressive and solenoidal components allows us to address the question of whether the core is {\it coherent}. This designation has been used to describe a core in which the velocity dispersion in its inner parts is dominated by the thermal speed, with turbulent motions being subsonic, and therefore subdominant in comparison to thermal motions as a source of pressure \citep{Barranco_Goodman98}, presumably due to turbulent dissipation at the core's center.

However, our numerical simulation does not support this picture. Figure \ref{fig:vcvs} shows the dispersion (standard deviation) of the solenoidal (blue lines) and compressive (red lines) components of the velocity, together with the total nonthermal velocity dispersion (green lines) as a function of radial distance from the center for various temporal snapshots. It is seen that the dispersion of the solenoidal component (which is purely turbulent) maintains a nearly uniform, {\it transonic} value throughout the whole radial extent of the core, and for the entire duration of the simulation. Instead, the compressive component increases continuously over time, as the infall speeds increase. So, in our core it is not that the (solenoidal) turbulence decreases at the center, but rather that 
it remains roughly constant (albeit it in fact increases slightly at late times in the inner regions), maintaining a transonic value. Interestingly, the solenoidal velocity component thus behaves in a nearly ``isothermal'' way, in the sense that its dispersion remains approximately constant.

As mentioned in Sec.\ \ref{sec:grav_amp}, this ``pseudo isothermal'' behavior differs from the approach to equipartition reported in the nonmagnetic case by \citet{RobG12} for externally-imposed contraction rates, and from the ``pseudo-virial'' behavior reported by \citet{GV20} in a self-consistent simulation of turbulent collapse. In the setup of the latter authors, the energy injection rate to the turbulence increases with time (as the infall speeds increase), and the kinetic energy in the turbulent component approaches roughly half that in the infall component.

As also discussed in Sec.\ \ref{sec:grav_amp}, the apparent discrepancy between those previous results and the behavior of the solenoidal component in our simulations can be understood if the solenoidal modes comprise only a moderate fraction of the total turbulent motions, with a significant part of the latter being in compressible, non-infall, turbulent motions. In this case, the solenoidal dispersion shown in Fig.\ \ref{fig:vcvs} constitutes only a lower limit to the total turbulent velocity dispersion. This can be seen, for example, in the inner regions where the red curve is below the blue curve in the frames corresponding to times $t = 0.8125 \tff$ and $t = 0.9375 \tff$ in Fig.\ \ref{fig:vcvs}. It is important to note, also, that this solenoidal component is clearly {\it not} dissipated towards the center, but instead is continuously driven by the collapse, as manifested by its nearly constant level over time. 

In view of the above, we can conclude that the decrease of the {\it total} turbulent velocity dispersion in the inner parts, as illustrated by the green lines in all the panels of Fig.\ \ref{fig:vcvs} for $t > 0$ is due to the decrease of the infall speed towards the center in the prestellar case, since the compressible turbulent speed appears to be mostly ``locked'' to (i.e., is a fixed fraction of) the infall speed. On the other hand, the solenoidal component, albeit maintaining a nearly radially uniform amplitude, is in most cases a small fraction of the total turbulent velocity dispersion. In the few cases where it is not, the total velocity dispersion departs from the infall speed.



\section{Summary and conclusions} \label{sec:concls}

In this paper, we have investigated several properties of the MHD turbulence generated by the gravitational collapse of a nearly spherical core seeded with slightly subsonic, solenoidal initial turbulence. We performed this analysis by decomposing the velocity field in its compressive and solenoidal components. Our results are as follows:

\begin{itemize}

\item In spite of the simulation being strongly magnetically supercritical, the collapse still proceeds significantly anisotropically, forming first a dense sheet that then collapses along its larger dimensions. The magnetic field in the central parts of the core is amplified by roughly two orders of magnitude due to the collapse.

\item The collapse amplifies the turbulent motions, but mostly in the compressible modes. The solenoidal modes remain almost at the initial level, although this means that they do not decay, either.

\item The amplitude of the solenoidal motions is roughly uniform with radius at all times, and roughly constant in time, indicating that no decrease of the turbulence occurs towards the center nor over time, as would be expected in the scenario that the turbulence decays inwards \citep[leaving behind a {\it coherent} core;] [] {Barranco_Goodman98} and over time, allowing the collapse of the core when turbulent support is lost. Nevertheless, the {\it total} velocity, including the dominant infall component, does decrease inwards during the prestellar stage investigated in this work, in agreement with the theoretical prediction from analytical spherical collapse calculations for the prestellar stage \citep{WS85}.

\item The distribution of the simulation cells in the $B$-$n$ space shows a clear superposition of the fast, slow, and Alfv\'en modes predicted by PV03, in addition to the effect of gravitational contraction: at low densities, a wide range range of $B$ values exists at a given density. This can be understood as a consequence of the slow and fast MHD modes scaling differently with density, as $B = a - bn$ for the slow mode (with $a$ and $b$ constants), and as $B \propto n$ for the fast mode. On the other hand, at high densities, the magnetic field scales as $n^\alpha$, with $1/2 < \alpha < 2/3$, indicating a domination of gravitational compression with either spherical or planar geometry.

\item As predicted by \citet{Gomez+21}, the mass-to-magnetic flux ratio ($M/\phi$) measured out to a certain radius scales (increases) with radius at all times, and, furthermore, the slope of the $M/\phi$ profile becomes shallower as time increases. In general, the outer regions of the core are magnetically supercritical and the inner regions are subcritical, in agreement with observations by, e.g., \citet{Crut09} at the scale of the transition from dense molecular core to its envelope, and by \citet{ChiLi22} at the interface between molecular clouds and their surrounding cold atomic gas.\footnote{In this regard, it is important to point out that the apparent subcriticality of the inner regions of a globally supercritical structure is a scale-free phenomenon that is expected to occur whenever a density fluctuation arises within a globally magnetically supercritical medium.} Moreover, at any given (fixed) radius $R$, $M/\phi$ increases with time. This is due to the accretion of material preferentially along field lines onto the region inside $R$.

\end{itemize}

We conclude that the MHD turbulence driven by gravitational contraction shares many of the features characterizing standard-driven turbulence, and moreover, that the driving by the collapse generates preferentially compressible components of the turbulence, while the solenoidal components remain at roughly the initial amplitude, although without decaying, either. The latter result implies that the turbulence in a collapsing core does not undergo a transition to coherence by dissipation, but rather the decrease in the linewidth at the inner, denser parts is due to the inwards decrease of the infall speed during the prestellar stage of the collapse. Finally, we have confirmed our earlier theoretical predictions that the mass-to-flux ratio is not expected to remain constant in time nor uniform in radius when the cores are defined by density thresholds.




\section*{Acknowledgements}
EVS acknowledges financial support from UNAM-PAPIIT grant IG100223. YH and AL acknowledge the support of NASA ATP AAH7546, NSF grants AST 2307840, and ALMA SOSPADA-016. Financial support for this work was provided by NASA through award 09\_0231 issued by the Universities Space Research Association, Inc. (USRA).

\section*{Data availability}
The data generated for this article will be shared upon reasonable request to the corresponding author.

\bibliographystyle{mnras}
\bibliography{xu}

\appendix

\section{Resolution Study} \label{sec:convergence}

As mentioned in Sec.\ \ref{sec:num_sim}, the larger number of cells per Jeans length ($j_{\rm r}$) used in the MHD run mhdturb\_08 (with maximum refinement level $\ell = 8$; see Table \ref{table:run_par}) requires a higher resolution level $\ell$ in order to preserve the value of the maximum resolved density. Conversely, at a fixed value of $\ell$, the maximum resolved density is lower at larger $j_{\rm r}$. Therefore, we have performed a magnetic simulation with $\ell = 10$ in order to test whether our results are significantly modified at higher resolution. 

\begin{figure*}
\centering
	\includegraphics[width=1.0\linewidth]{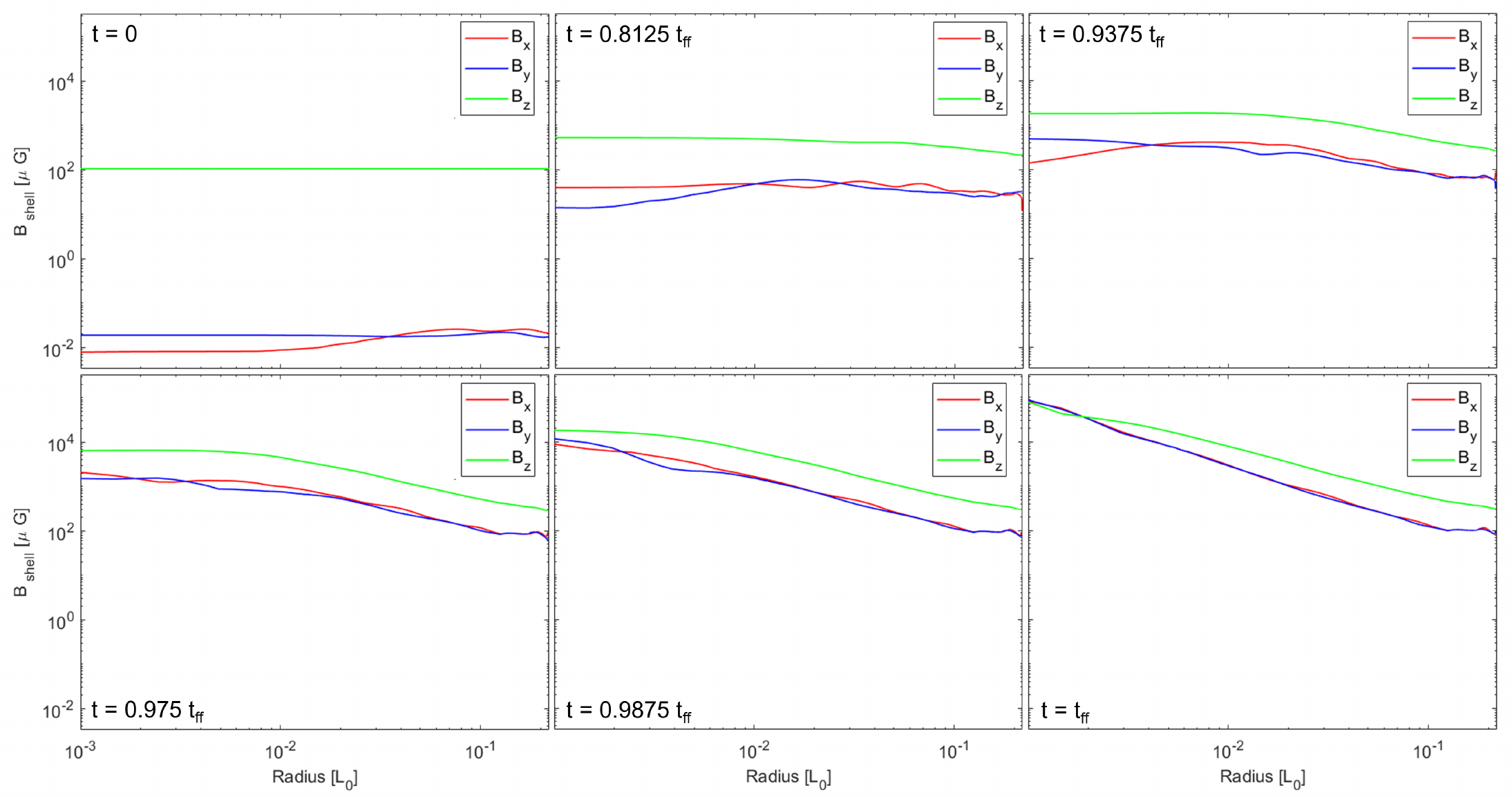}
    \caption{Like Fig.\ \ref{fig:Bshell}, but for the high-resolution run {\it mhdturb\_10}. No significant qualitative change in the radial profile of the magnetic field components is seen at the increased resolution.}
    \label{fig:b_512}
\end{figure*}

Figure \ref{fig:b_512} presents the same plots as Fig.\ \ref{fig:Bshell}, showing the radial profile of the shell-averaged magnitude of the magnetic field at various times, for the high-resolution run {\it mhdturb\_10}. No significant change in the radial profile of the magnetic field components is seen in comparison to run {\it mhdturb\_08}.



\begin{figure*}
\centering
	\includegraphics[width=1.0\linewidth]{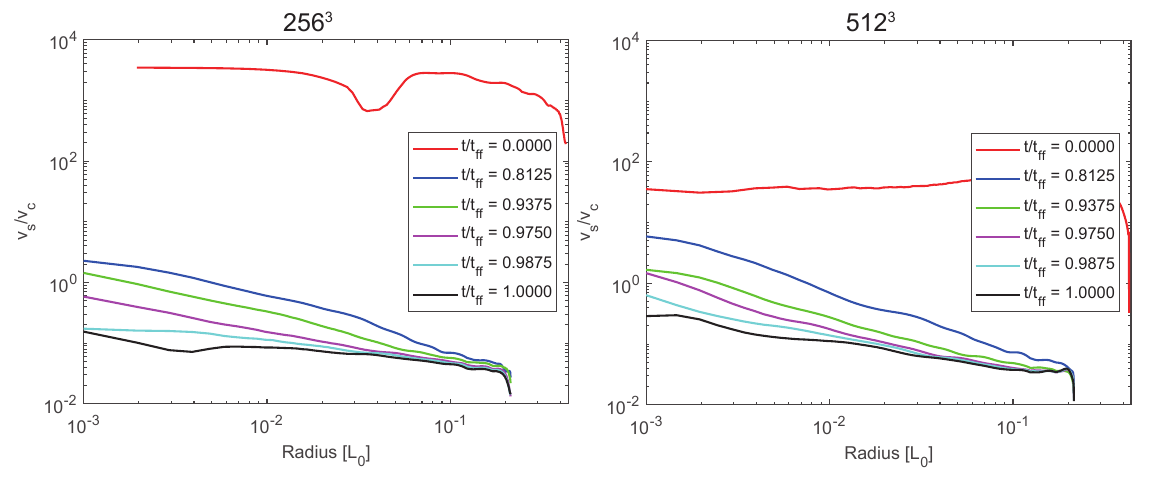}
    \caption{Radial profile of the ratio of the RMS values of the solenoidal to compressible parts of the velocity for the low- (left) and high-resolution (right) simulations.}
    \label{fig:v_ratio_512}
\end{figure*}

Also, Fig.~\ref{fig:v_ratio_512} compares the radial profiles of the ratio of the solenoidal to compressible components in the low- and high-resolution runs. We notice the ratio at $t=0$ in the high-resolution run is approximately one order of magnitude lower than that of the low-resolution runs. We found this comes from the fact that the high-resolution runs at $t=0$ have more significant compressive velocity, while their total velocity and solenoidal velocity are at similar levels as the low-resolution runs. Since all the simulations are driven with only solenoidal modes during the pre-gravity driving stage, we speculate that the higher compressible fraction is due to a more efficient transfer from the solenoidal to the compressible modes. Nevertheless, the ratio is still large at all radii at $t=0$ and, more importantly, no significant qualitative difference is seen between the different resolutions at the later snapshots, suggesting that the results reported in the body of the paper, based on run {\it mhdturb\_08} are not affected by any possible lack of resolution.

\label{lastpage}
\end{document}